\documentclass[12pt]{article}
\usepackage{graphicx}
\usepackage{enumerate}
\usepackage{natbib}
\usepackage{url} 

\usepackage{amsmath, amssymb}
\usepackage[colorlinks,citecolor=blue,urlcolor=blue]{hyperref}
\usepackage{enumitem}

\usepackage{multirow}
\usepackage{shortcuts}
\usepackage{graphicx}
\usepackage{nicefrac}
\usepackage{subcaption}
\usepackage{booktabs}
\usepackage{todonotes}
\usepackage[flushleft]{threeparttable}
\usepackage{bm}
\usepackage{bbm}
\usepackage[flushleft]{threeparttable}
\usepackage{float}
\usepackage{extarrows} 
\usepackage{adjustbox}

\usepackage{amsthm}

\usepackage{accents}
\newlength{\dhatheight}

\numberwithin{equation}{section}
\newtheorem{assumption}{Assumption}[section]

\newcommand{\blind}{1}

\begin{document}

\def\spacingset#1{\renewcommand{\baselinestretch}%
{#1}\small\normalsize} \spacingset{1}


\if1\blind
{
  \title{\bf Robust weights that optimally balance confounders for estimating marginal hazard ratios}
  \author{Michele Santacatterina\thanks{
    The author gratefully acknowledge Xiaojie Mao and Angela Zhou for valuable comments and help on this paper}\hspace{.2cm}\\
    Department of Biostatistics and Bioinformatics, \\ George Washington University}
  \maketitle
} \fi

\if0\blind
{
  \bigskip
  \bigskip
  \bigskip
  \begin{center}
    {\LARGE\bf Title}
\end{center}
  \medskip
} \fi

\bigskip
\begin{abstract}
Covariate balance is crucial in obtaining unbiased estimates of treatment effects in observational studies. Methods based on inverse probability weights have been widely used to estimate treatment effects with observational data.  Machine learning techniques have been proposed to estimate propensity scores. These techniques however target accuracy instead of covariate balance. Methods that target covariate balance have been successfully proposed and largely applied to  estimate  treatment  effects on continuous outcomes.  However,  in  many  medical  and  epidemiological  applications,  the interest lies in estimating treatment effects on time-to-an-event outcomes.With this type of data, one of the most common estimands of interest is the marginal hazard ratio of the Cox proportional hazard model. In this paper, we start by presenting robust orthogonality weights (ROW), a set of weights obtained by solving a quadratic constrained optimization problem that maximizes precision while constraining covariate balance defined as the sample correlation between confounders and treatment. By doing so, ROW optimally deal with both binary and continuous treatments. We then evaluate the performance of the proposed weights in estimating marginal hazard ratios of binary and continuous treatments with time-to-event outcomes in a simulation study. We finally apply ROW on the evaluation of the effect of hormone therapy on time to coronary heart disease and on the effect of red meat consumption on time to colon cancer among 24,069 postmenopausal women enrolled in the Women's Health Initiative observational study. 
\end{abstract}

\noindent%
\vfill

\newpage
\spacingset{1.45} 

\section{Introduction}
\label{sec:intro}

Covariate balance is crucial in obtaining unbiased estimates of treatment effects in observational studies. Weighted methods based on Inverse Probability Weights (IPW) have been used to estimate the effect of a treatment on an outcome using observational data. IPW are constructed as the inverse of the probability of a unit being assigned to a treatment conditional to pre-treatment covariates, \textit{i.e.}, the propensity score \citep{rosenbaum1983central}. Despite their wide use, IPW-based methods tediously rely on the correct specification of the propensity score model, which violations lead to biased estimates, and on the positivity assumption \citep{imbens2015causal}, which \textit{practical} violations \citep{petersen2012diagnosing} lead to extreme weights and erroneous inferences \citep{robins1995analysis,scharfstein1999adjusting,robins2007comment,kang2007demystifying}. 
Machine learning techniques, like the Super Learner \citep{van2007super}, have been proposed to improve propensity score estimation in the case of model misspecification \citep{lee2010improving,pirracchio2015improving}. These techniques however target accuracy instead of covariate balance. 


Methods that mitigate model misspecification while targeting covariate balance have been proposed. Among others, \cite{imai2014covariate} proposed and extended \citep{fong2018covariate} Covariate Balancing Propensity Score (CBPS), which uses generalized method of moments to estimate the logistic regression model that optimally balances covariates.  \cite{zubizarreta2015stable} proposed Stable Balancing Weights (SBW), a set of weights with minimal sample variance that satisfy a list of approximate moment matching conditions. \cite{hainmueller2012entropy} presented entropy balancing weights obtained by minimizing the entropy of the weights while satisfying balance conditions. The literature of covariate balance is extensive and many other methods have been developed in the recent years \citep[among others]{kallus2019optimal,kallus2019kernel,2018arXiv180601083K, hirshberg2019minimax,hirshberg2017augmented,zhao2017entropy,zhao2019covariate,wong2017kernel,visconti2018handling,zubizarreta2014matching,li2018balancing,king2017balance,tubbicke2020entropy,vegetabile2020nonparametric,wu2018matching,josey2020framework, yiu2018covariate} 

Although methods that target covariate balance have been shown to be robust to either model misspecification, practical positivity violation or both, these methodologies have been mainly developed and applied to estimate treatment effects on continuous outcomes. However, in many medical and epidemiological applications, the interest lies in estimating the effect of a treatment effects on time-to-an-event outcomes. Examples include, the evaluation of the impact of hormone therapy on time to coronary heart disease (CHD) \citep{hulley1998randomized,manson2003estrogen}, and the impact of red meat consumption on time to colon cancer \citep{larsson2005red}. When estimating treatment effects with time-to-event data one of the most common causal estimands of interest is the marginal hazard ratio of the Cox proportional hazard model (although other estimands may be of interest such as the survival difference or the mean survival \citep{mao2018propensity}).


In this paper, based on the framework introduced by \cite{yiu2018covariate}, we start by presenting robust orthogonality weights (ROW),  which optimally and robustly balance covariates for estimating effects of binary and continuous treatments. We then evaluate its performance in estimating marginal hazard ratios of binary and continuous treatments with time-to-event outcomes. The proposed weights are obtained by solving a convex constrained quadratic optimization problem which minimize the sample variance of the weights, thus controlling for extreme weights and maximizing precision, while constraining the sample correlation between treatment and covariates, thus optimally balance covariates with respect to either binary or continuous treatments. Similar to IPW, matching, CBPS and SBW, the set of ROW is obtained without the use of the outcome, thus emulating randomization. 
As discussed in \cite{yiu2018covariate}, by minimizing the sample variance of the weights while controlling for a measure of balance, ROW is constructed in the spirit of SBW. However, differently from SBW, ROW can also be used to estimate effects of continuous treatments. Also, by using the sample correlation as a measure of balance, ROW can be seen as an extension of CBPS \citep{fong2018covariate} in which precision is maximized while satisfying a constraint on the covariate balance. 

Our contribution to this field of literature is to provide a set of weights of minimal variance that optimally balance confounders for estimating marginal hazard ratios of binary and continuous treatments with time-to-event data. In addition, we contribute to this literature, by providing a thorough comparison of the performance of several covariate balancing methods with time-to-event data in multiple simulation scenarios and in two case studies. Finally, we provide a \texttt{R} package for the computation of the weights available at \href{https://github.com/michelesantacatterina/ROW}{https://github.com/michelesantacatterina/ROW}. Code for simulations and case studies analyses is available at \href{https://github.com/michelesantacatterina/ROW-time-to-event}{https://github.com/michelesantacatterina/ROW-time-to-event}.


In the next Section we present ROW and provide some practical guidelines on their use. In Section \ref{sec:simu}, we evaluate the performance of ROW with respect to absolute bias, root mean squared error, covariate balance and computational time across levels of practical positivity violation, misspecification and censoring for both binary and continuous treatments. In Section \ref{sec:case-studies}, we apply ROW to  the  evaluation  of  the  effect  of  hormone  therapy  on time to first occurrence of coronary heart disease and the impact of red meat consumption on time to colon cancer onset using data from 24,069 postmenopausal women enrolled in the Women’s Health Initiative observational study \citep{study1998design}. We provide some concluding remarks in Section \ref{sec:conclusions}.


\section{Robust orthogonality weights}
\label{row}

In this section, we first introduce some notations and assumptions, we then present the optimization problem used to obtain robust orthogonality weights and finally discuss some practical guidelines. 

\subsection{Notations and assumptions}
\label{sec:notation_assumptions}

Suppose we have a simple random sample of size $n$ from a target population. For each unit $i=1,\dots, n$, suppose we observe a binary or continuous treatment $A_i$, a set of pre-treatment covariates (also referred to as confounders) $\mathbf{X}_i$, and the observed time to an event, $T_i$. In addition, let $C_i$ denote the $i^{th}$ individual's censoring time, $\Delta_i=\I{T_i<C_i}$, the complete-case indicator and $Y_i$ the observed event only if $\Delta_i=1$. We define the potential (counterfactual) follow-up time and response as $T_i(a)$ and $Y_i(a)$ respectively. Throughout this paper, in addition to consistency and non-interference \citep{imbens2015causal}, we assume 
\begin{assumption}
\textit{Strong ignorability}  $\lbrace T_i(a), Y_i(a) \rbrace \independent A_i | \mathbf{X}_i$,
\end{assumption}
\begin{assumption}
\textit{Noninformative censoring}\footnote{the survival time provides no information about the distribution of censoring, and vice-versa \cite{kalbfleisch2011statistical}},
\end{assumption}
\begin{assumption}
\textit{Positivity} $\phi(\mathbf{X}_i)=P(A_i=a | \mathbf{X}_i)>0$,
\end{assumption}
\noindent
where $\phi(\mathbf{X}_i)$ is the classic propensity score \citep{rosenbaum1983central} if the treatment is binary and the generalized propensity score as presented in \cite{hirano2004propensity} if the treatment is continuous, \ie  the conditional density of the treatment given covariates, $r(a,x)=f_{A|X}(a|x)$. The main causal estimand of interest is the marginal hazard ratio, \ie the $\theta$ parameter of the Cox proportional hazard model 
\begin{align*}
    \lambda ( t | A_i ) = \lambda_0(t) \exp \left( \theta A_i \right).
\end{align*}
To estimate $\theta$, standard practice suggests to use an outcome model such as the conditional Cox regression model $\lambda(t|A_i, \mathbf{X}_i)=\lambda_0(t) \exp \left( \theta A_i + \beta \mathbf{X}_i \right)$ (a conditional estimator), or an IPW-marginal Cox model, \ie an IPW-weighted Cox regression, regressing only the treatment on the time to the event, with propensity scores estimated using regression or machine learning techniques (see \cite{buchanan2014worth} for an applied example). These procedures provide biased and erroneous inferences when the outcome model or the propensity score model is misspecified or when the positivity assumption is practically violated. In addition, covariate balance is not targeted. In the next Section we introduce a convex quadratic constrained optimization problem to obtain robust orthogonality weights that target covariate balance and are robust to practical positivity violations and misspecification.



\subsection{A convex quadratic constrained optimization problem}

A general measure of covariate balance is the correlation between treatment and covariates. When the correlation is equal to zero, the covariates are uncorrelated from the treatment. Based on the framework of \cite{yiu2018covariate} and in the spirit of \cite{zubizarreta2015stable}, we propose to find weights with minimum variance while satisfying contraints on the sample correlation between covariates and the treatment under study. We therefore propose to obtain ROW by solving the following quadratic linearly constrained optimization problem,

\begin{eqnarray}
\underset{\textbf{w}}{\text{minimize}} \label{op1}
& & \| \textbf{w} - e_n \|_2^2 \\ \label{of1}
\text{subject to} 
\label{op2}
& & | \rho_k(\textbf{w}) | \leq \delta, \; k = 1, \ldots, m, \\ 
\label{op3}
& & e^\top \textbf{w} = 1, \\
& & \textbf{w} \geq 0,
\label{op4}
\end{eqnarray}
\noindent
where $\rho_k(\textbf{w}) = \sum_{i=1}^n w_i X^\ast_{i,k}A^\ast_{i,k}$ is the (weighted) mean of the products of the standardized covariates and treatment, \ie, the sample correlation, $X^\ast$ and $A^\ast$ are the scaled covariates and treatment variables, $m$ is the total number of covariates, $e$ is the unit vector, and $e_n$ is the unit vector divided by the sample size $n$ which, by construction, represent the mean of the weights. When a solution to optimization problem \eqref{op1}-\eqref{op4} exists, constraint \eqref{op2} guarantees that the correlation between treatment and covariates is at most equal to $\delta$ (for each covariate), and constraints \eqref{op3} and \eqref{op4} guarantee that the weights sum up to 1 and are positive, respectively.  

\begin{figure}
\begin{center}
\includegraphics[scale=1]{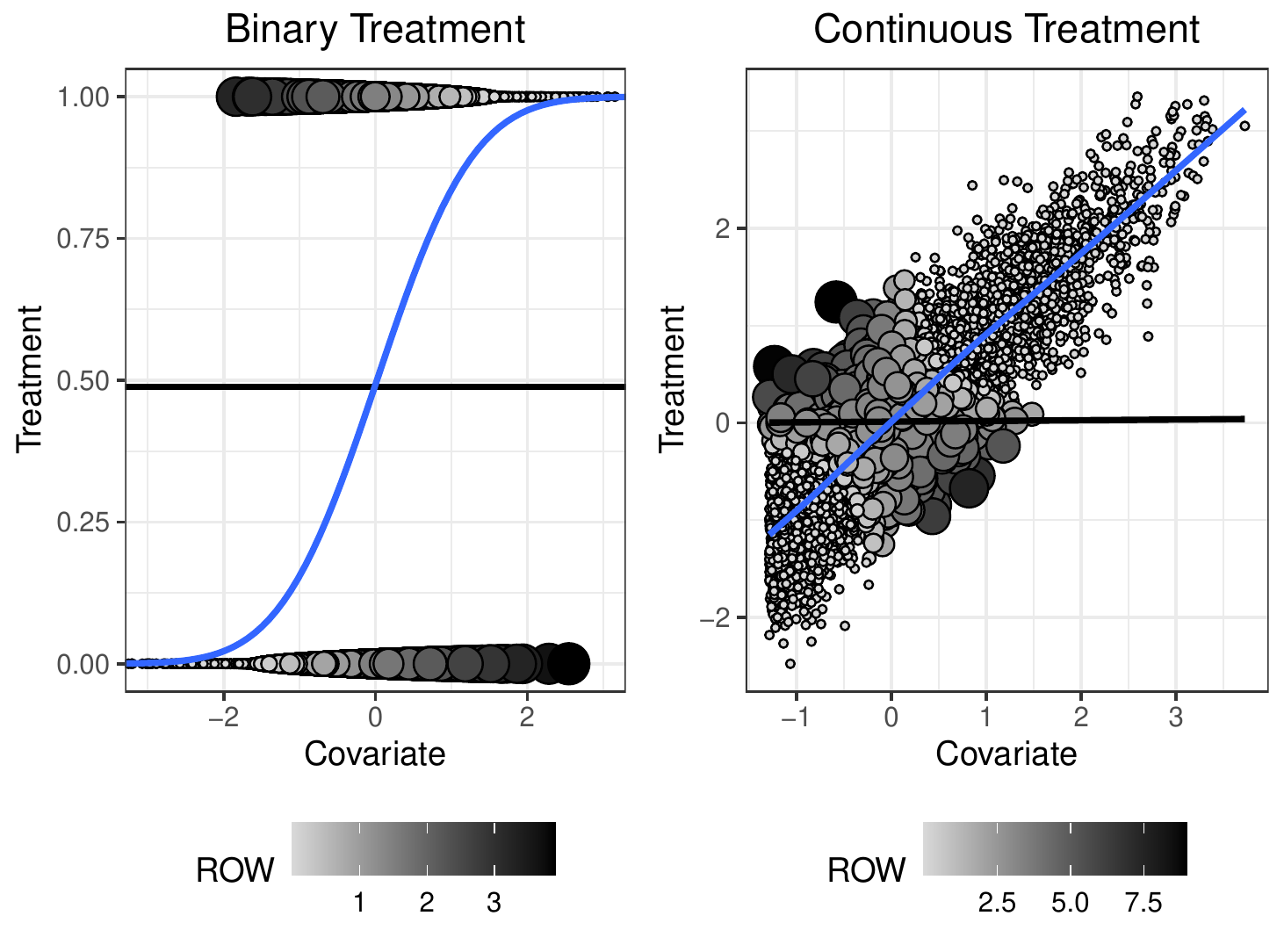}
\end{center}
\caption{\footnotesize Graphical representation of ROW balancing a covariate (x-axis) across a binary (left panel) and continuous (right panel) treatments (y-axis). Blue lines represent the true relationships between the binary (a probit model) and the continuous (simple regression with normal errors and positive coefficient) treatment. Black lines represent the relationship between treatments and covariate after weighting for ROW. Size and color of the circles represent the individual weight assigned (the larger/darker the higher). 
\label{fig_explain_w} }
\end{figure}

What does the set of ROW achieve? In Figure \ref{fig_explain_w}, we provide a simple scenario in which a binary treatment (left panel of Figure \ref{fig_explain_w}) and a continuous treatment (right panel) are generated as a function of a continuous covariate (x-axis), using a probit model and a simple regression model with normal errors, respectively. Blue lines represent the true relationship between the covariate and the treatment. For instance, in the continuous treatment scenario, the covariate has a positive impact on the treatment, \ie the regression coefficient is positive. The black lines represent the values of the regression coefficients after weighting by ROW (we first computed ROW by solving optimization problem \eqref{op1}-\eqref{op4} and then plug the obtained weights into a weighted probit and ordinary regression estimators, regressing the covariate on the treatment). Under the binary treatment scenario, the weighted coefficient equals 0.5, while it equals 0 in the continuous treatment scenario for each value of the covariate. In summary, these figures show that ROW orthogonalize covariate and treatment variables thus eliminating associations between them. By doing so, as shown in our simulations in Section \ref{sec:simu} and in our case studies in Section \ref{sec:case-studies} ROW maximize covariate balance. Figure \ref{fig_deps} in Section \ref{sm:sec:additional results} of the Supplementary Material shows that ROW orthogonalize covariates and treatment variables also when the true relationship between them is nonlinear quadratic, nonlinear cubic, nonlinear without correlation, sinusoidal, and when the treatment is right- and left-skewed. In addition Figure \ref{fig_deps} shows that under independence between treatment and a covariate, ROW result in almost uniform weights. 

\subsection{Practical guidelines}
In this section, we provide some practical guidelines on the choice of the parameter $\delta$, and on standard error estimation for the marginal hazard ratio.

\begin{figure}
\begin{center}
\includegraphics[scale=1]{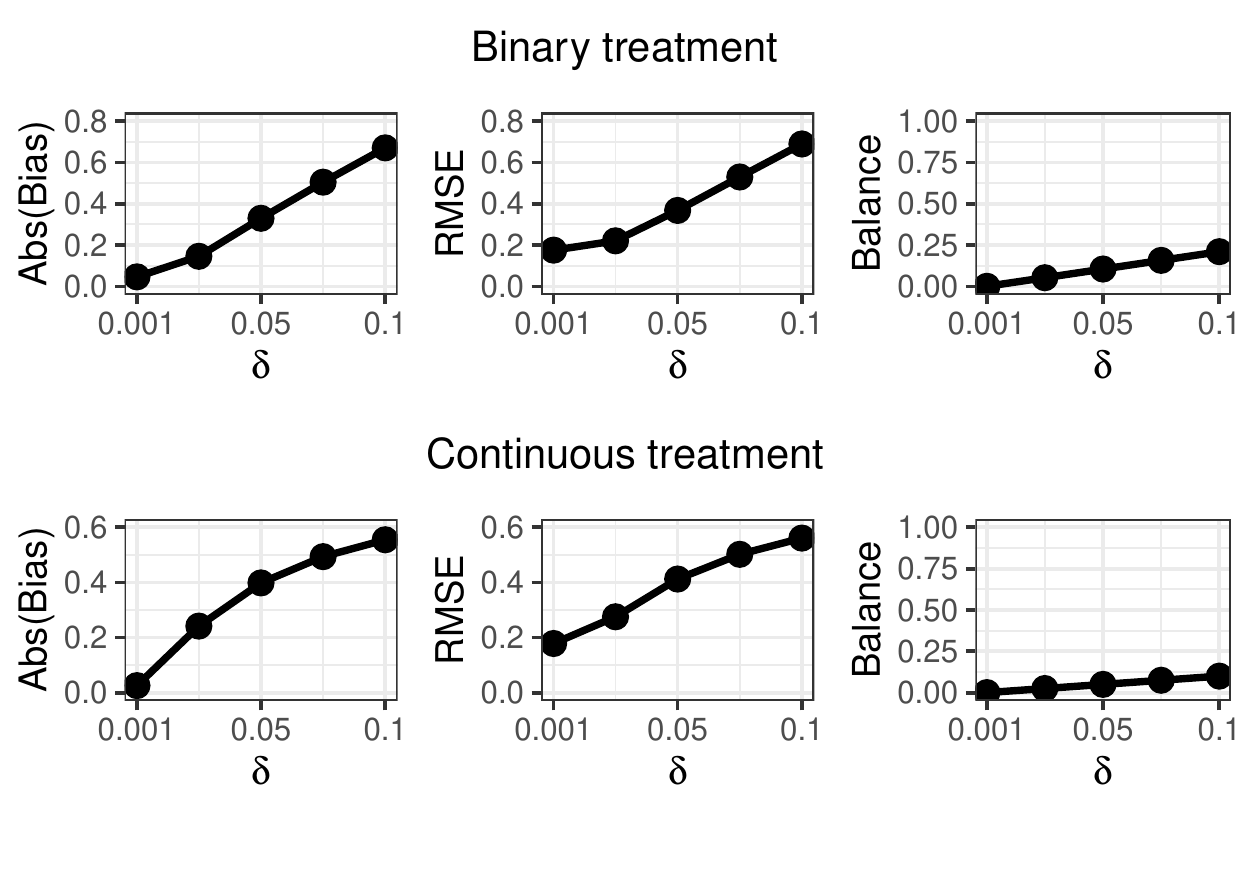}
\end{center}
\caption{\footnotesize Absolute bias (left panels), root mean squared error (RMSE) (middle panels), absolute standardized mean difference (Balance)(top right panel), absolute correlation (bottom right panel), for the binary (top panels) and continuous (bottom panels) treatment across levels of the balance constraint $\delta$ (eq. \eqref{op1}).
\label{fig_delta} }
\end{figure}

Optimization problem \eqref{op1}-\eqref{op4} depends on the parameter $\delta$. This parameter set the upper bound for the absolute value of the sample correlation between treatment and covariates. Smaller values of $\delta$ induce smaller correlation, thus inducing higher balance and consequently lower bias. Figure \ref{fig_delta} provides a graphical representation of the impact of $\delta$ on bias, mean squared error and balance. Precisely, it shows absolute bias (left panels), root mean square error (middle panels) of the marginal hazard ratio estimated using a Cox regression model weighted by ROW, and mean covariate balance across four continuous covariates (right panels) for binary (upper panels) and continuous treatments (lower panels), across levels of the parameter $\delta$, set equal to $0.001, 0.025, 0.05, 0.075$, and $0.1$ (Simulations details are provided in section \ref{sec:simu}). For the binary treatment, we considered the absolute standardized mean difference as a measure of balance while for the continuous treatment we considered the absolute correlation between treatment and covariates. Lower values of $\delta$ guarantee the lowest absolute bias, and balance. In addition, since optimization problem \eqref{op1}-\eqref{op4} constraints imbalance while controlling precision by minimizing the variance of the weights, lower values of $\delta$ also guarantee minimal root mean squared error for both types of treatments. While, theoretically, $\delta$ could be set equal to 0, to avoid numerical instabilities, we suggest to set $\delta=0.001$ or $\delta=0.0001$.

As suggested by other authors \citep{hernan2001marginal,robins2000marginal}, we suggest to use the robust ``sandwich'' estimator  \citep{freedman2006so} for the estimation of the standard error. When computational resources are not limited, we also suggest to use bootstrap \citep{davison1997bootstrap}. We provide a comparison between the naive (the inverse of the observed Fisher information of $\hat{\theta}$, the estimated marginal hazard ratio of the Cox hazard model introduced in Section \ref{sec:notation_assumptions}), robust and bootstrap standard errors in Section \ref{sm:simu_se_cove} of the Supplementary Material. To compute the bootstrap standard error, we used non-parametric bootstrap with normal approximation confidence intervals \citep{davison1997bootstrap}.

Similarly to \cite{zubizarreta2015stable} and \cite{santacatterina2018optimal,santacatterina2019optimal}, Lagrange multipliers can be used to evaluate the impact that a small decrease in the parameter $\delta$ would cause in the objective function \eqref{of1}. We refer to Section 3.3 of \cite{zubizarreta2015stable} and Section 3 and 3.1 of \cite{santacatterina2018optimal} and \citep{santacatterina2019optimal}, respectively for detail. 

Many solvers are available to solve constrained convex quadratic optimization problems. We suggest using \texttt{Gurobi} \citep{gurobi}.

Once the set of weights has been computed, to obtain marginal hazard ratios, we suggest using a weighted Cox proportional hazard model, regressing the treatment on the time-to-event on the ROW-weighted population. We also suggest to follow standard practice to evaluate the validity of the proportional hazard assumption.


\section{Simulations}
\label{sec:simu}
In this section we evaluate the performance of ROW with respect to, absolute bias, root mean square error, covariate balance and computational time, across levels of practical positivity violations, misspecification and censoring when estimating the marginal hazard ratio with a binary and continuous treatment. In summary, ROW performs well across all of the considered scenarios.

\subsection{Setup}
\label{simu_setup}
We considered a sample size of $N=1,000$. We computed the expected survival time $t$ by following the inverse probability method based on the Weibull distribution \citep{bender2005generating},
\begin{align*}
    T_i = \left( - \frac{\log(u)}{\psi \exp \left( \theta A_i + \textbf{X}_i^\top \mathbf{\beta}  \right)} \right)^{\frac{1}{\rho}},
\end{align*}
\noindent
where $u \sim \text{Unif}(0,1)$, $\psi=0.01$ (the scale parameter of the Weibull distribution), $\theta=0.2$, $X_1 \sim \text{N}(0.1,1), X_2 \sim \text{N}(0.1,1), X_3 \sim \text{logN}(0,0.5), X_4 \sim 5\text{Beta}(3,1)$ $X_5$ is a random sample with replacement of size 4 with probabilities $0.35,0.25, 0.05, 0.35$ respectively, and $X_6 \sim \text{Binom}(0.25)$, $\mathbf{\beta}=(0,1,0,1.4,1.4,1)$, and $\rho=1$ (the shape parameter of the Weibull distribution). With the choice of these six covariates we wanted to reflect real-world populations in which the time to an event depends on binary, categorical and continuous (normal and non-normal) covariates.   We generated the censoring times $C_i$ using an exponential distribution, \ie $C_i \sim \text{Exp}(\epsilon)$, with values for the rate parameter $\epsilon$ described in section \ref{simu_cens}. Finally, we obtained the observed (censored) survival times by taking the minimum between $T_i$ and $C_i$. The causal estimand of interest is the marginal hazard ratio (HR), $HR=\exp \theta =  1.22$. We provide detailed information on how we generated $A_i$ in the following Section. 

\subsubsection{Estimating HR for binary and continuous treatments}
\label{simu_bin_cont}
To evaluate the performance of ROW, we considered estimating the marginal hazard ratio under two scenarios: binary treatments and continuous treatments. We refer to the first scenario as the \textit{binary} scenario and the second as the \textit{continuous} scenario. In the binary scenario, we considered $A_i \sim \text{Binom}(\pi(\textbf{X}_i))$, where $\pi(\textbf{X}_i) = \left( 1+ \exp ( \gamma \left( \frac{\kappa}{\gamma} -\textbf{X}_i^\top \mathbf{e} \right) ) \right)^{-1}$ and $\kappa=n^{-1} \sum_{i=1}^n \left( \textbf{X}_i^\top \gamma \right)$. In the continuous scenario, we considered $A_i \sim \mu(\textbf{X}_i) + \text{logNorm}(0,\eta^2)$, where $\mu(\textbf{X}_i) =  - \kappa + \textbf{X}_i^\top \mathbf{e}$. 
We provide detailed information on the parameters $\gamma$ and $\eta$ in the following Section. 

\subsubsection{Estimating HR across levels of practical positivity violations}
\label{simu_ppv}
To evaluate the performance of ROW across levels of practical positivity violations for the binary scenario, we considered five values for $\gamma$, from 0.1 to 2. We refer to $\gamma=0.1$ as weak violation, $\gamma=1$ as moderate violation and to $\gamma=2$ as strong violation. The propensity score ranged from 0.13 to 0.61 under weak violation, 0.05 to 0.97 under moderate violation and 0.001 to 0.995 under strong violation (average of min/max propensities). In the continuous scenario, we considered five different values (0,0.1,0.5,0.6,0.7,0.9) for the parameter $\eta$ of the log-Normal distribution, \ie the (log) standard deviation of the random variable which higher values generate a more right-skewed distribution. We refer to $\eta=0$ as weak violation, $\eta=0.6$ as moderate violation and $\eta=0.9$ as strong violation.  The rate parameter $\epsilon$ for the censoring level was set equal to 0.01 (low percentage of censored observation - see Section \ref{simu_cens} for details) and we considered correct specification of the treatment model (see Section \ref{simu_miss} for details).

\subsubsection{Estimating HR across levels of misspecification}
\label{simu_miss}
We evaluated the performance of ROW across levels of misspecification. Specifically, for the binary scenario, we generated $Z_1=(X_1+0.5)^2$, $Z_2=((X_1X_2)/5 + 1)^2$, $Z_3=\exp(X_3/2)$, and $Z_4=X_4(1+\exp(X_3)) + 1$. For the continuous scenario we generated $Z_1=\exp(X_1/2)$, $Z_2=X_2(1+\exp(X_1)) + 1$, $Z_3=(X_1X_3/25 + 0.2)^3$, and $Z_4 = 2*\log(\abs{X_4})$. We then considered a convex combination between the correct variables $(X_1,X_2,X_3,X_4)$ and the misspecified variables $(Z_1,Z_2,Z_3,Z_4)$, \eg $X_1=X_1(1-\tau) + Z_1\tau$, and let $\tau$ vary from 0 to 1 (0,0.25,0.5,0.75,1). We refer to $\tau=0$ as null misspecification, $\tau=0.5$ as moderate misspecification and $\tau=1$ as strong misspecification.  The rate parameter $\epsilon$ for the censoring level was set equal to 0.01 (low percentage of censored observation - see Section \ref{simu_cens} for details) and we considered moderate pratical positivity violation (see Section \ref{simu_ppv} for details).

\subsubsection{Estimating HR across levels of censoring}
\label{simu_cens}
We also evaluated the performance of ROW across levels of censoring. We considered five values (0,1,10,100,1000) for the rate parameter $\epsilon$ of the exponential distribution used to generate the censoring times, resulting in 1, 7, 25, 52 and 78 percent of censored observations in the binary scenario and in 2, 10, 27, 53 and 76 percent of censored observations in the continuous scenario. We considered correct specification (see Section \ref{simu_miss} for details) and moderate pratical positivity violation (see Section \ref{simu_ppv} for details).

\subsubsection{Methods comparison}
\label{simu_meth}

In addition to the standard of practice methods such as the conditional Cox proportional hazard model and IPW-Cox regression, we consider methods that 1) use only the information of the treatment and covariates and not the outcome, 2) target covariate balance in some way and 3) their \texttt{R} implementation is readily available.  

For the binary scenario, we compared ROW with IPW-Cox regression, where propensity scores were estimated by using SuperLearner (IPW) with the following library of algorithms: logistic regression model with only main effects, logistic regression with main effects and interactions, lasso-penalized logistic regression, random forest, bayesian logistic regression and extreme gradient boosting classifier; Balance SuperLearner (BalSL) as described in \cite{pirracchio2018balance} with the same library of algorithms as for IPW; and by using boosted logistic regression (GBM) (with absolute standardized mean difference as stopping criteria, interaction depth equal to 3, number of trees equal to 10,000, shrinkage equal to 0.01 and bag fraction equal to 1) \citep{mccaffrey2004propensity}. In addition, we compared ROW  with  Propensity Score Matching (PSM), with propensity scores obtained as for IPW \citep{sekhon2008multivariate}; Covariate Balancing Propensity Score (CBPS) containing only the covariate balancing conditions (exact identification); entropy balancing weights (EBAL) \citep{hainmueller2012entropy}; stable balance weights (SBW) \citep{zubizarreta2015stable} where we selected the degree of approximate covariates balance by following the tuning algorithm presented in \cite{wang2020minimal}, (we chose the grid of values for the tuning algorithm equal to 0.0001, 0.001, 0.002, 0.005, 0.01, 0.02, 0.05, and 0.1); outcome model (OM), a Cox proportional hazard model regressing confounders and treatment on the time to event; and a Cox regression model conditioning only on the binary treatment (naive). 

For the continuous scenario, we compared ROW  with IPW-Cox regression, where propensity scores were estimated by using SuperLearner (IPW) with the following library of algorithms: linear regression with only main terms, linear regression with main terms and interactions, lasso-penalized linear regression, random forest, Bayesian linear regression, local polynomial regression, and extreme gradient boosting regressor; Balance SuperLearner (BalSL) with the same library of algorithms as of IPW; and by using gradient boosted regression (GBM) (with  Pearson  correlation  between  covariates  and  treatment  as  stopping  criteria, interaction depth equal to 4,  number of trees equal to 20,000,  shrinkage equal to 0.0005 and bag fraction equal to 1). The final IPW weights were obtained assuming Normal conditional density of the treatment as suggested by \cite{robins2000marginal}. In addition, we compared ROW with Covariate Balancing Propensity Score (CBPS) containing only the covariate balancing conditions; non-parametric CBPS (npCBPS); outcome model (OM), a Cox proportional hazard model conditioned on confounders and treatment; and a Cox regression model conditioning only on the continuous treatment (naive).

\subsection{Additional simulations}

In addition to the simulations presented in section \ref{simu_setup}, we provide additional simulations to evaluate the impact of 1) practical positivity violation, misspecification and censoring on coverage of the 95\% confidence interval; 2) sample sizes and 3) number of covariates included in the analysis, on absolute bias, root mean squared error, balance and computational time in seconds.  We provide a summary of the results in section \ref{simu_res_additional} and detailed results in Section \ref{sm:simu} of the Supplementary Material. 

\subsection{Results}

\subsubsection{Binary treatment}

\begin{figure}
\begin{center}
\includegraphics[scale=.65]{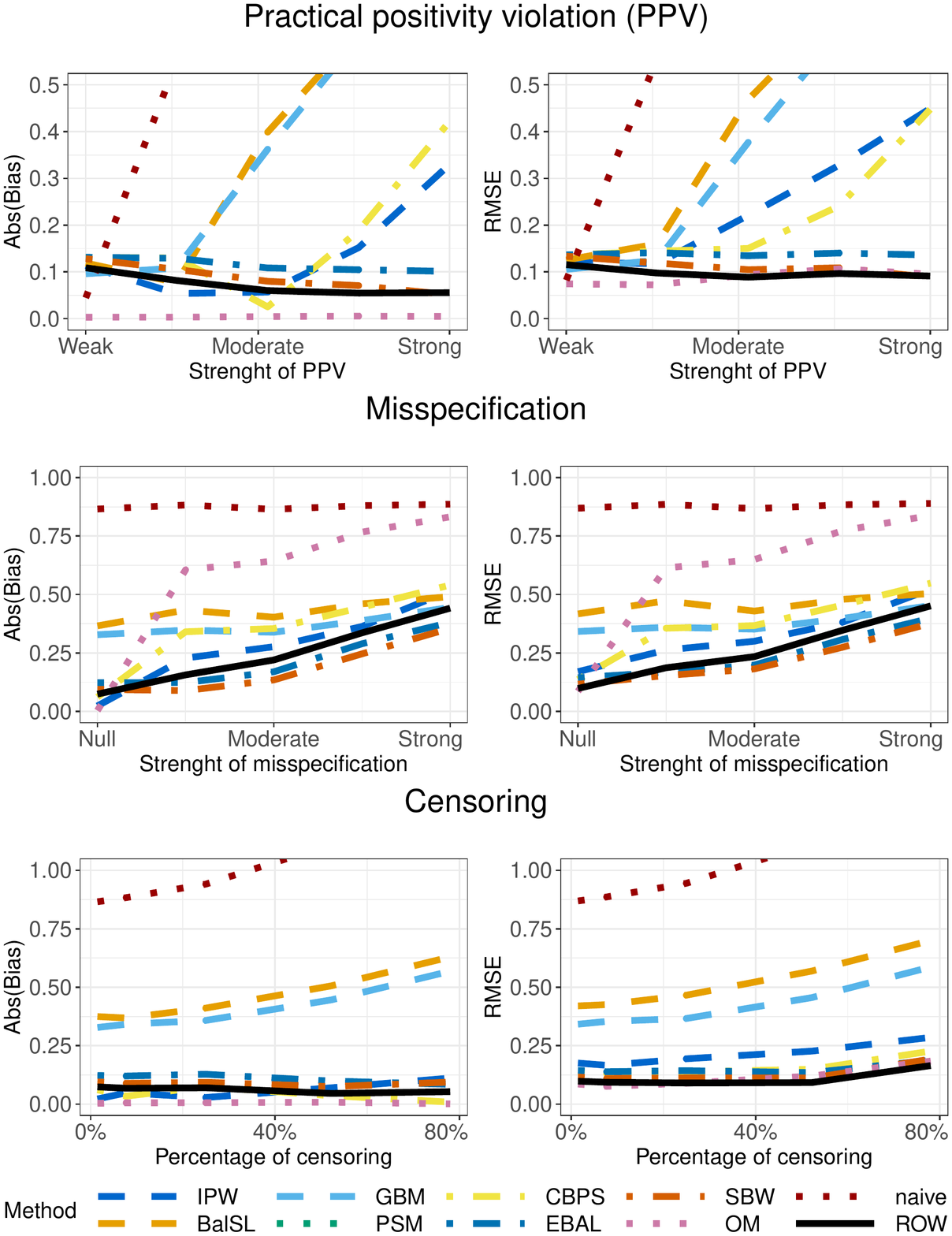}
\end{center}
\caption{\footnotesize \textit{Binary treatment}: Absolute bias (left panels) and root mean squared error (RMSE) (right panels) across levels of practical positivity violations (top panels), misspecification (middle panels) and censoring(bottom panels) when estimating the marginal hazard ratio of a binary treatment.  ROW: Robust Optimal Weights; IPW: Inverse Probability Weighting;  GBM:  propensity  scores  were  estimated  with  Gradient  Boosting Machine; CBPS: Covariate Balancing Propensity Score; SBW: Stable Balancing Weights; Naive:  Cox proportional hazard model including only the treatment; BalSL: Balance  SuperLearner;  PSM: Propensity Score Matching (values outside figures); EBAL: Entropy Balancing; OM: (outcome model) Cox proportional hazard model including confounders and treatment. 
\label{fig_bias_mse_binary} }
\end{figure}

In summary, ROW performed well across all simulation scenarios. Figure \ref{fig_bias_mse_binary} shows absolute bias (left panels) and root mean squared error (RMSE) (right panels) across levels of practical positivity violations (top panels), misspecification (middle panels) and censoring (bottom panels) when estimating the marginal hazard ratio of a binary treatment. ROW (black-solid line), EBAL (dark-blue dashed-dotted) and SBW (red dashed-dotted) performed well overall. IPW (blue dashed), BalSL (orange dashed), and GBM (light-blue dashed) performed moderately well across all levels of misspecification and censoring but showed higher bias and RMSE for moderate and strong violation of the positivity assumption. These results suggest that flexible models for the estimation of the propensity scores, may mitigate possible misspecification, but lead to erroneous inferences in the presence of lack of covariate overlap. Similar results were obtain for CBPS (yellow dotted-dashed). OM (purple dotted) outperformed all other methods across levels of positivity violation but performed worse across levels of misspecification. Contrary to previous literature \citep{austin2013performance}, we found that all methods outperformed PSM (green dotted; values are outside figures).


\subsubsection{Continuous treatment}

\begin{figure}
\begin{center}
\includegraphics[scale=.65]{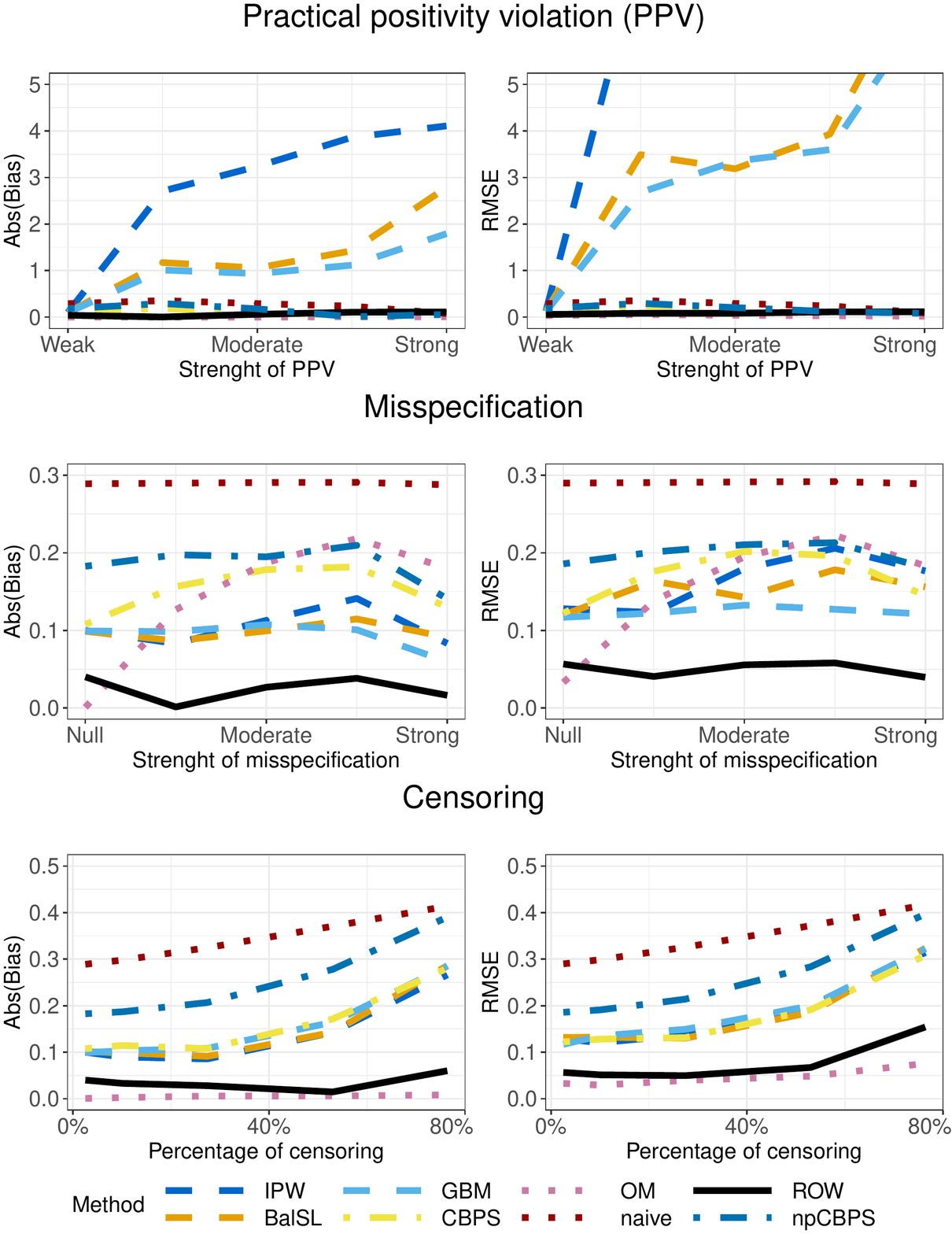}
\end{center}
\caption{\footnotesize \textit{Continuous treatment}: Absolute bias (left panels) and root mean squared error (RMSE) (right panels) across levels of practical positivity violations (top panels), misspecification (middle panels) and censoring(bottom panels) when estimating the marginal hazard ratio of a binary treatment.  ROW: Robust Optimal Weights; IPW: Inverse Probability Weighting (propensity scores were estimated with SuperLearner;  GBM:  propensity  scores  were  estimated  with  Gradient  Boosting Machine; OM: (outcome model) Cox proportional hazard model including confounders and treatment; BalSL: Balance  SuperLearner;  CBPS: Covariate Balancing Propensity Score;  Naive:  Cox proportional hazard model including only the treatment;  npCBPS: non-parametric CBPS.
\label{fig_bias_mse_cont} }
\end{figure}

 ROW performed well across all simulation scenarios, especially across levels of misspecification. Figure \ref{fig_bias_mse_cont} shows absolute bias (left panels) and root mean squared error (RMSE) (right panels) across levels of practical positivity violations (top panels), misspecification (middle panels) and censoring (bottom panels) when estimating the hazard ratio of a continuous treatment. As for the binary treatment scenario, IPW (blue dashed), BalSL (orange dashed), and GBM (light-blue dashed) performed well across levels of misspecification and censoring but were outperformed by all other methods across levels of practical positivity violation. CBPS (yellow dotted-dashed) and npCBPS (dark-blue dotted-dashed) performed well across levels of positivity violations and censoring and performed similarly to IPW, BalSL and GBM across levels of misspecification. OM performed worse across levels of misspecification. 

\subsubsection{Summary of additional simulations' results}
\label{simu_res_additional}

By using either the robust or bootstrap standard error, ROW achieved desirable coverage under weak and moderate practical positivity violation and misspecification for both, binary and continuous treatment (Figures \ref{fig_se_cove} and \ref{fig_se_cove_c} in the Supplementary Material).  Under strong practical positivity violation and misspecification, ROW showed undercoverage due to increased bias, regardless of the use of the robust, bootstrap or naive standard error. ROW achieved desirable levels under all censoring levels. Absolute bias and RMSE decreased while increasing the sample size (top panels of Figures \ref{fig_ss_b} and \ref{fig_ss_c}  in the Supplementary Material). ROW achieved low balance (below 0.05 standardized absolute mean difference across confounders for the binary treatment scenario and below 0.025 absolute correlation across confounders for the continuous treatment scenario) across all sample sizes considered. Although balance was kept at low levels, absolute bias and RMSE increased while increasing the number of covariates (top panels and left bottom panels of Figures \ref{fig_num_c_b} and \ref{fig_num_c_c} in the Supplementary Material). This can be explained by the fact that we generated the data by setting the coefficient for each confounders equal to 1, regardless of the number of confounders considered, thus leading to a strongly confounded model when the number of confounders increased (detailed description of the simulation setting is provided in Section \ref{sm:simu_num_cove} of the Supplementary Material). Since we plan to apply ROW using observational data from medical registries, we were especially interested in the computational burden needed to find a solution for larger sample sizes and for an increased number of confounders. We found that for relatively large sample sizes, \eg $n=10,000$ the solver could find a solution in a few seconds (bottom right panels of Figures \ref{fig_ss_b} and \ref{fig_ss_c} in the Supplementary Material). In our case-study presented in Section \ref{sm:sec:case-studies} in which we balanced 36 confounders on a population of $n=24,069$ individuals, the solver found a solution in about 12 seconds.

\section{Case studies}
\label{sec:case-studies}

In this section, we apply ROW to the evaluation of the effect of hormone therapy on coronary heart disease and the impact of red meat consumption on colon cancer using data from the Women's Health Initiative observational study \citep{study1998design}. 

\subsection{The effect of hormone therapy on coronary heart disease}
\label{ht_chd}

The Women's Health Initiative (WHI) is a long-term study of postmenopausal women that focuses on best strategies for the prevention and treatment of heart diseases, breast and colon cancers and other chronic diseases. WHI is composed of a randomized clinical trial and an observational study. The WHI trial aimed at evaluating the health benefits and risks of hormone therapy when taken for chronic disease prevention among predominantly healthy postmenopausal women \citep{writing2002risks}.  Specifically, one of the trial's component evaluated the impact of estrogen plus progestin therapy on the risk of coronary heart disease (CHD). Prior to this trial, large observational studies suggested that postmenopausal hormone users had a reduced risk of CHD \citep{stampfer1991estrogen,grady1992hormone, sidney1997myocardial,psaty1994risk}. In contrary, results from the WHI trial suggested an increased risk of CHD \citep{writing2002risks}. Precisely, in the original trial, the Authors found a statistically significant estimated marginal hazard ratio equal to 1.29 (1.02-1.63).

In this section, we aim at evaluating the effect of estrogen plus progestin therapy on time to CHD among postmenopausal women aged 50-79 years using data from the WHI observational study (September 1993-September 2010). Following \citet{hernan2008observational}, we first mimic the design of the WHI trial as closely as possible in the WHI observational study. We then apply ROW to control for the non-randomization of the treatment. We also compare the estimated marginal hazard ratio obtained by using ROW, with those obtained by using the methods presented in section \ref{simu_meth}.

\subsubsection{Study population}

We considered the target study population of postmenopausal women who in the WHI observational study had reported no use of estrogen therapy, progesterone therapy or their combination during 2-year prior the enrollment in the study. Baseline was defined as first follow-up visit and women were followed from baseline to diagnosis of CHD, loss to follow-up, death, or September 30, 2010, whichever occurred first. Out of the 93,676 women comprising the original WHI observational study 37,080 used any hormone therapy in the 2-year before the enrollment of the study while 30,960 lacked information about the number of days since enrollment, and 1,567 lacked information on time since menopause. The final study population was comprised of 24,069 women. 

We considered the following 34 confounders: multivitamine without minerals use (yes, no), multivitamine with minerals use (yes, no), ethnicity (White, Black, Hispanic, Native American, Asian/Pacific Islander, Unknown), number of pregnancies (7 categories), bilateral oophorectomy (yes, no), age at menopause (numeric), breast cancer ever (yes, no),  colon cancer ever  (yes, no), endometrial cancer ever  (yes, no), skin cancer ever  (yes, no), melanoma cancer ever (yes, no), other cancer past 10 years  (yes, no), deep vein thrombosis ever (yes, no), stroke ever (yes, no), myocardial infarction ever (yes, no), diabetes ever (yes, no), high cholesterol requiring pills ever (yes, no), osteoporosis ever  (yes, no), cardiovascular disease ever  (yes, no), coronary artery bypass graft  (yes, no), atrial fibrillation ever  (yes, no), aortic aneurysm ever  (yes, no), angina  (yes, no), hip fracture age 55 or older  (yes, no), smoked at least 100 cigarettes ever (yes, no), alcohol intake (non drinker, past drinker, less than 1 drink per month, less than 1 drink per week, 1 to 7 drinks per week, 7+ drinks per week), fruits med serv/day (numeric), vegetables med serv/day (numeric), dietary energy (kcal), systolic blood pressure (numeric), diastolic blood pressure (numeric), body mass index (numeric), education (11 categories), income (10 categories).

 Time since menopause has been recognized as an important factor for the risks and benefits of hormone therapy on CHD \citep{carrasquilla2015association,carrasquilla2017postmenopausal}. We therefore evaluated the impact of estrogen plus progestin therapy on time to CHD by conducting a stratified analysis on three categories of time since menopause: 0-10 years, 11-20 years and 20+ years.

To impute missing values of the aforementioned confounders we used multiple imputation by chained equations \citep{buuren2010mice}. marginal hazard ratio estimates were computed using the imputed dataset.

\subsubsection{Models setup}

We obtained ROW by solving optimization problem \eqref{op1}-\eqref{op4} setting $\delta=0.0001$ in constraint \eqref{op2} (for each confounder). To obtain the set of ROW we used the \texttt{R} interface of \texttt{Gurobi}. We compared ROW with IPW in which we estimated the propensity score by using SuperLearner with the following library of algorithms: logistic  regression  model  with  only  main  effects,  lasso-penalized  logistic  regression, and random  forest; BalSL with the same library of algorithms as that of IPW; GBM (with mean absolute standardized mean difference as stopping method, interaction depth equal to 3, number of trees equal to 10,000, shrinkage equal to 0.01 and bag fraction equal to 1); PSM with propensity scores estimated as for IPW and BalSL; CBPS containing only the covariate balancing conditions (exact identification); EBAL and SBW with balance tolerance in standard deviation set equal to 0.0001 (with the sample size of our dataset, the tuning algorithm presented in \cite{wang2020minimal} used to choose the degree of approximate covariates balance for SBW considerably increased the computational burden of SBW and therefore was not performed), OM: outcome modelling, a Cox proportional hazard model including confounders and treatment; and Naive: Cox proportional hazard model including only the treatment. To obtain the set of SBW we also used \texttt{Gurobi}.  Once the sets of weights were obtained, we plugged the weights into a weighted Cox regression regressing the treatment (estrogen plus progestin therapy) on the time to CHD. We computed robust (sandwich) standard errors.

\subsubsection{Results}

\begin{figure}
\begin{center}
\includegraphics[scale=.55]{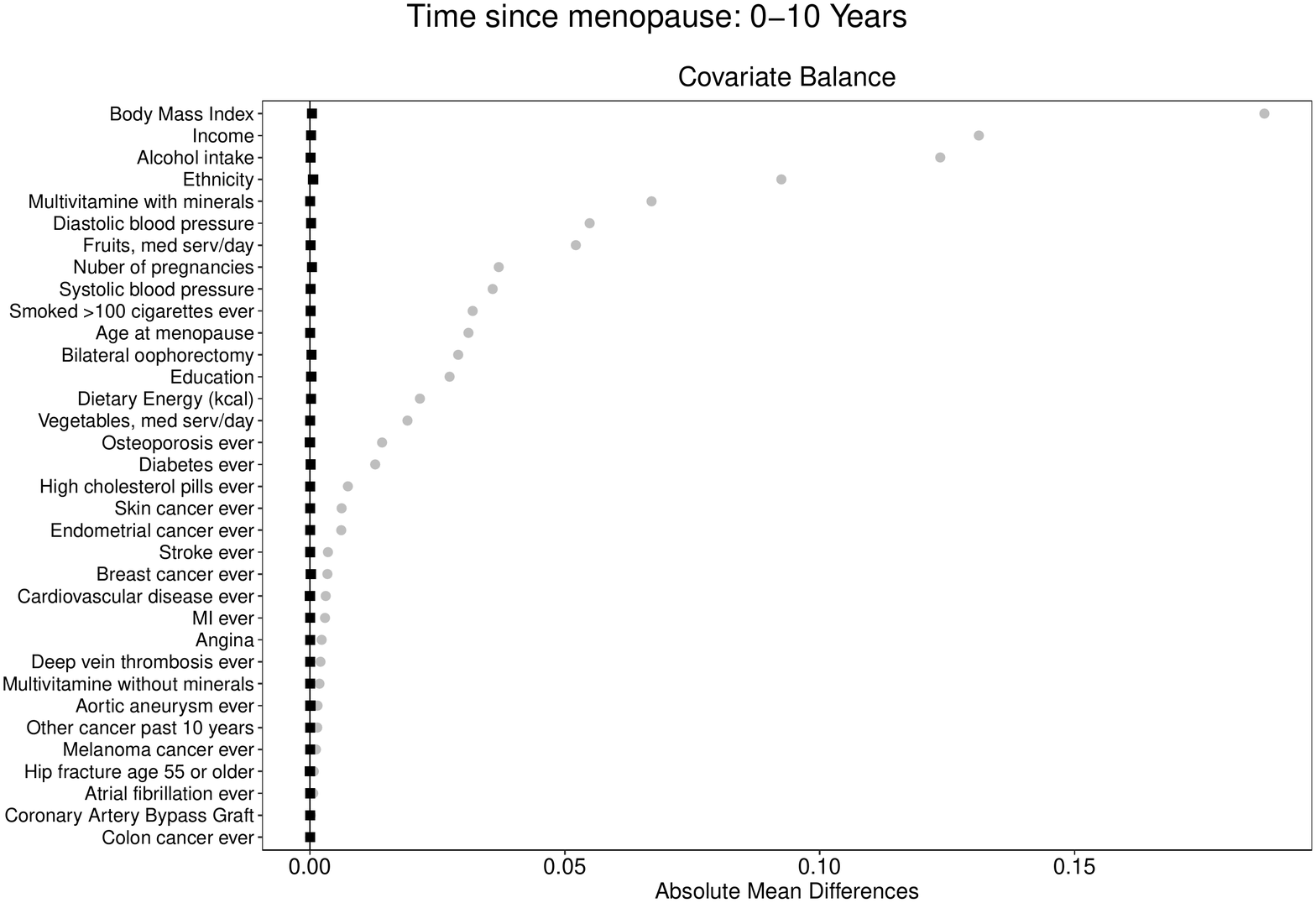}
\end{center}
\caption{Adjusted (weighted by ROW) (black squares) and unadjusted (grey dots) absolute standardized mean differences between confounders and treatment (estrogen plus progestin).  
\label{fig_1} }
\end{figure}

\begin{figure}
\begin{center}
\includegraphics[scale=.55]{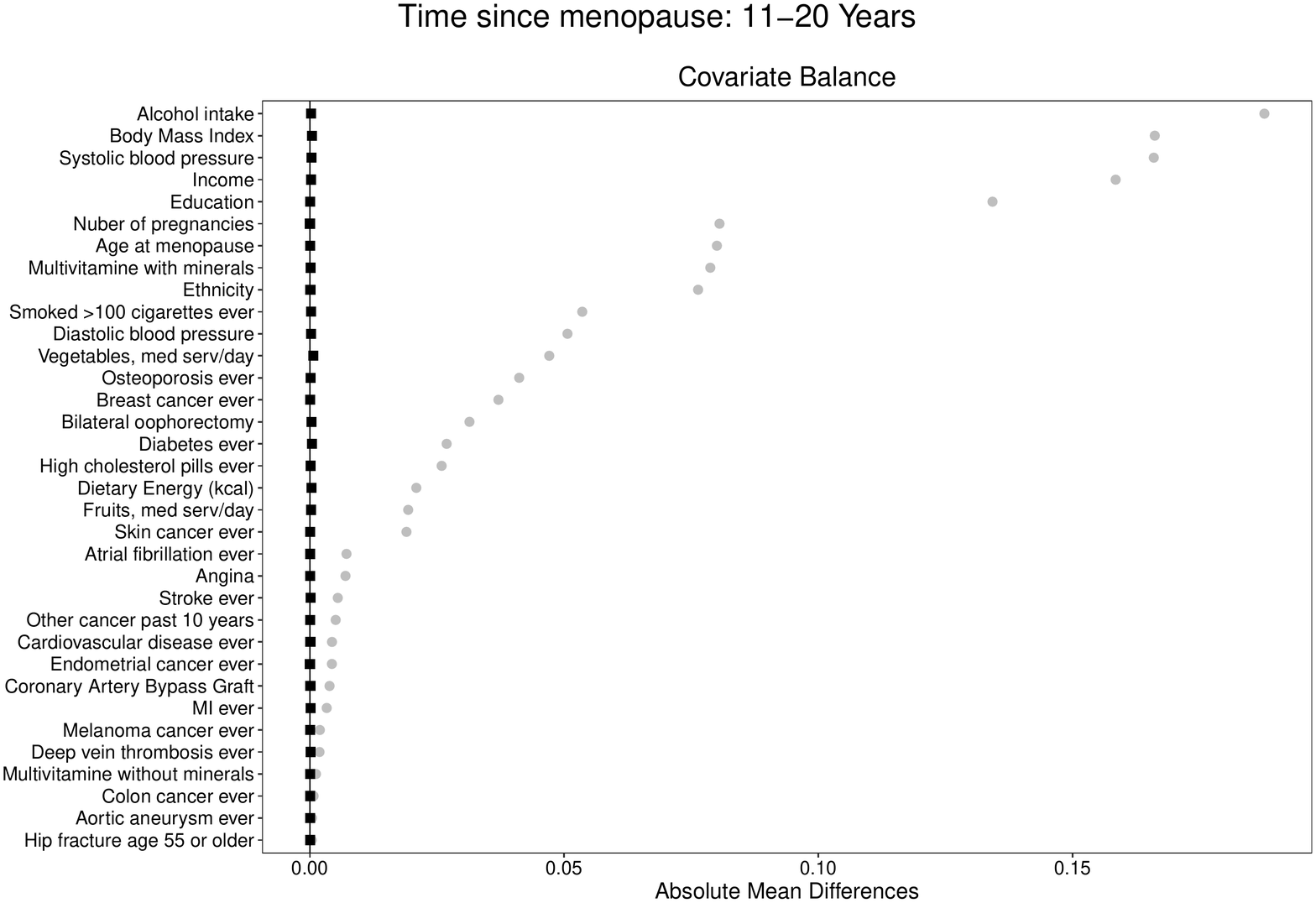}
\end{center}
\caption{Adjusted (weighted by ROW) (black squares) and unadjusted (grey dots) absolute standardized mean differences between confounders and treatment (estrogen plus progestin). 
\label{fig_2} }
\end{figure}

\begin{figure}
\begin{center}
\includegraphics[scale=.55]{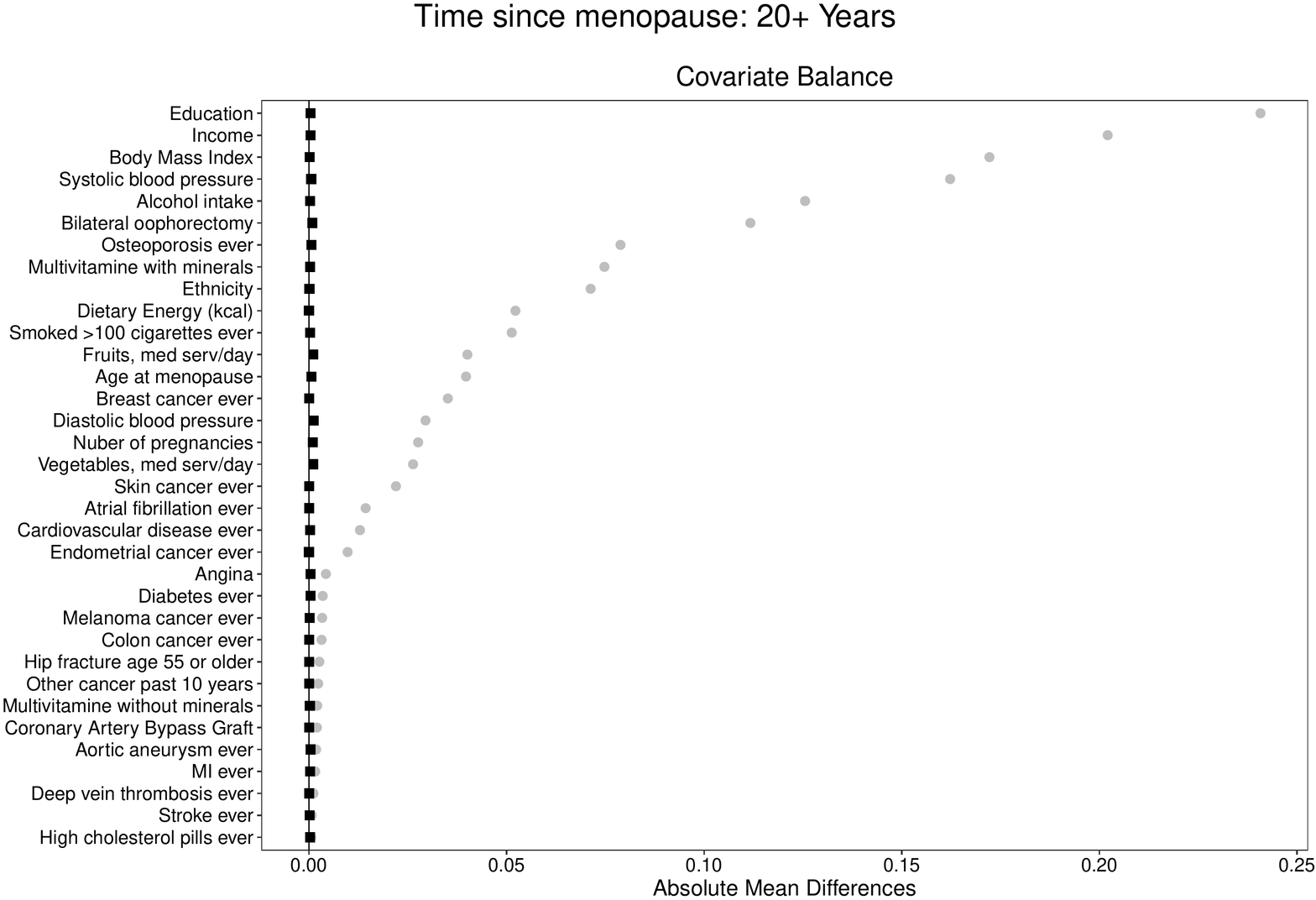}
\end{center}
\caption{ Adjusted (weighted by ROW) (black squares) and unadjusted (grey dots) absolute standardized mean differences between confounders and treatment (estrogen plus progestin).
\label{fig_3} }
\end{figure}

\textbf{Covariate balance.} Figures \ref{fig_1}-\ref{fig_3} show absolute mean differences (standardized for continuous variables, raw for binary variables) between each of the aforementioned 34 confounders and the binary treatment estrogen plus progestin therapy versus no therapy, before (grey dots) and after (black squares) weighting for ROW.  ROW successfully balance all the confounders across the three strata of time since menopause. Table \ref{table_b2} shows the minimum, median and maximum absolute mean difference (standardized for continuous variables and raw for binary variables) between confounders and treatment across 34 confounders and across categories of time since menopause for each of the considered methods. ROW, SBW, EBAL, and CBPS achieved the lowest maximum absolute mean difference across confounders, with ROW obtaining the lowest for strata 0-10 years and 20+ years. Methods that target precision instead of balance, such as IPW with SuperLearner, balanced confounders worse than those methods that target covariate balance, such as ROW, SBW, CBPS or EBAL. SBW could not find a solution when setting the balance tolerance equal to 0.0001 (as described in the previous Section), 0.001, 0.01, and 0.1 in the 0-10 years and 20+ years category. We consequently do not report any results for SBW in these categories.  Table \ref{table_b3} in the Supplementary Material, shows the computational time in seconds needed to obtain a solution across strata of time since menopause for each of the considered method. The computational time required by ROW was significantly lower than that needed by IPW, BalSL, GBP, SBW and PSM. Computational time slightly increased with sample size.

\noindent
\textbf{Outcome analysis.} Table \ref{table_b1} shows marginal hazard ratio estimates and 95\% confidence intervals (computed using the robust standard error) for the effect of estrogen plus progestin therapy on time to CHD, and the robust standard error for each method and across categories of time since menopause. Most of the methods lead to similar results. Proportional hazards tests \citep{grambsch1994proportional} resulted in p-values greater than 0.05 for all models.  Figure \ref{fig_km} in the Supplementary Material shows the (adjusted) Kaplan-Meier curves weighted by ROW for estrogen plus progestin (HT) across categories of time since menopause. 

Based on assumptions 2.1-2.3 of Section \ref{sec:notation_assumptions}, the simulation results presented in Section \ref{sec:simu}, and on the covariate balance performance of ROW presented in Figures \ref{fig_1}-\ref{fig_3} and Table \ref{table_b2}, we conclude that estrogen plus progestin therapy has no statistically significant effect on time to CHD among $n=24,069$ postmenopausal  women aged 50-79 years enrolled in the WHI observational study (September 1993-September 2010), across three categories of time since menopause, \ie, $\hat{HR}$ and $95\%$ confidence intervals equal to  1.26 (0.70;2.28),  1.33 (0.87;2.02), and 0.79 (0.45;1.39), for 0-10, 11-20 and 20+ years since menopause, respectively.

\begin{table}[]
\centering
\begin{threeparttable}
\caption{Minimum, median and maximum absolute mean difference (standardized  for continuous variables, raw for binary variables) between confounders and treatment across 34 confounders and across categories of time since menopause (0-10; 11-20; 20+ years), WHI observational study, $n=24,069$, 1993-2010. \label{table_b2}}
\setlength\tabcolsep{1.5pt} 
\begin{tabular}{cccccccccc}
\hline
\textbf{}                           & \multicolumn{9}{c}{\textbf{Time since menopause}}                                                                                                                                                            \\ \cline{2-10} 
\textbf{}                           & \multicolumn{3}{c}{\textbf{0-10 Years}}                            & \multicolumn{3}{c}{\textbf{11-20 Years}}                           & \multicolumn{3}{c}{\textbf{20+ Years}}                             \\
\textbf{}                           & \multicolumn{3}{c}{$n=6,661$}                            & \multicolumn{3}{c}{$n=9,592$}                           & \multicolumn{3}{c}{$n=7,816$}                             \\
\textbf{}                           & \multicolumn{3}{c}{\textbf{Abs Mean Diff}}                & \multicolumn{3}{c}{\textbf{Abs Mean Diff}}                & \multicolumn{3}{c}{\textbf{Abs Mean Diff}}                \\
\textbf{}                           & \textbf{Min}         & \textbf{Median}      & \textbf{Max}         & \textbf{Min}         & \textbf{Median}      & \textbf{Max}         & \textbf{Min}         & \textbf{Median}      & \textbf{Max}         \\ \cline{2-10}
\multicolumn{1}{l}{\textbf{Method}} & \multicolumn{1}{l}{} & \multicolumn{1}{l}{} & \multicolumn{1}{l}{} & \multicolumn{1}{l}{} & \multicolumn{1}{l}{} & \multicolumn{1}{l}{} & \multicolumn{1}{l}{} & \multicolumn{1}{l}{} & \multicolumn{1}{l}{} \\ \cline{1-1}
\textbf{ROW}                        & \textless{}0.0001    & \textless{}0.0001    & 0.0006               & \textless{}0.0001    & 0.0001               & 0.0007               & \textless{}0.0001    & 0.0003               & 0.0012               \\
\textbf{IPW}                        & 0.0001               & 0.0071               & 0.1239               & 0.0003               & 0.0124               & 0.0736               & 0.0002               & 0.0111               & 0.0609               \\
\textbf{BalSL}                      & 0.0002               & 0.0032               & 0.057                & 0.0007               & 0.0065               & 0.1195               & 0.0007               & 0.0065               & 0.1195               \\
\textbf{GBM}                        & \textless{}0.0001    & 0.0035               & 0.0537               & \textless{}0.0001    & 0.0054               & 0.0704               & 0.0007               & 0.0075               & 0.0689               \\
\textbf{CBPS}                       & \textless{}0.0001    & 0.0002               & 0.0022               & \textless{}0.0001    & 0.0002               & 0.0007               & \textless{}0.0001    & 0.0004               & 0.0022               \\
\textbf{SBW}                       & -     & -    & -                           & \textless{}0.0001    & \textless{}0.0001      & 0.0002               & -    & -               & -               \\
\textbf{EBAL}                       & \textless{}0.0001    & \textless{}0.0001    & 0.0012               & \textless{}0.0001    & \textless{}0.0001    & \textless{}0.0001    & \textless{}0.0001    & \textless{}0.0001    & 0.0017               \\
\textbf{PSM}                        & 0.0006               & 0.0185               & 0.3357               & 0.0002               & 0.0162               & 0.2411               & 0.0007               & 0.0227               & 0.1601               \\
\textbf{Naive}                      & \textless{}0.0001    & 0.0101               & 0.1872               & 0.0004               & 0.0234               & 0.1877               & 0.0006               & 0.0242               & 0.2408               \\ \hline
\end{tabular}
    \begin{tablenotes}
      \small
      \item \textit{First column}: Method implemented, ROW: Robust Optimal Weights; IPW: Inverse Probability Weighting (propensity scores were estimated with SuperLearner with linear  regression  model  with  only  main  effects,  and random  forest in the library of algorithms); BalSL: Balance SuperLearner; GBM: propensity scores were estimated with Gradient Boosting Machine; CBPS: Covariate Balancing Propensity Score; SBW: Stable Balancing Weights; EBAL: Entropy Balancing; PSM: Propensity Score Matching;  OM: (outcome model) Cox  proportional  hazard model including confounders and treatment; Naive:  Cox  proportional  hazard model including only the treatment. \textit{Second, Fourth and Sixth columns}: Marginal hazard ratio and 95\% confidence interval (computed using robust standard error). \textit{Third, Fifth and Seventh columns}: robust standard error.
    \end{tablenotes}
  \end{threeparttable}
\end{table}

\begin{table}[]
\centering
\begin{threeparttable}
\caption{Marginal hazard ratio estimate and 95\% confidence intervals of the effect of estrogen plus progestin therapy on time to CHD among  postmenopausal  women between 50 and 79 across categories of time since menopause (0-10; 11-20; 20+ years), WHI observational study, $n=24,069$, 1993-2010. \label{table_b1}}
\begin{tabular}{ccccccc}
\hline
\textbf{}       & \multicolumn{6}{c}{\textbf{Time since menopause}}                                                                                       \\ \cline{2-7} 
\textbf{}       & \textbf{0-10 Years}  & \textbf{}            & \textbf{11-20 Years} & \textbf{}            & \textbf{20+ Years}   & \textbf{}            \\
\textbf{}       & $n=6,661$  & \textbf{}            & $n=9,592$ & \textbf{}            & $n=7,816$   & \textbf{}            \\
\textbf{}       & $\hat{HR}$ ($95\%$ CI)          & \textbf{SE}          & $\hat{HR}$ ($95\%$ CI)          & \textbf{SE}          & $\hat{HR}$ ($95\%$ CI)          & \textbf{SE}          \\ \cline{2-7} 
\textbf{Method} & \multicolumn{1}{l}{} & \multicolumn{1}{l}{} & \multicolumn{1}{l}{} & \multicolumn{1}{l}{} & \multicolumn{1}{l}{} & \multicolumn{1}{l}{} \\ \cline{1-1}
\textbf{ROW}    & 1.26 (0.70;2.28)      & 0.301                & 1.33 (0.87;2.02)     & 0.216                & 0.79 (0.45;1.39)     & 0.285                \\
\textbf{IPW}    & 1.23 (0.69;2.17)     & 0.292                & 1.24 (0.83;1.84)     & 0.202                & 0.79 (0.46;1.36)     & 0.275                \\
\textbf{BalSL}  & 1.25 (0.70;2.23)      & 0.295                & 1.26 (0.85;1.89)     & 0.204                & 0.82 (0.48;1.38)     & 0.269                \\
\textbf{GBM}    & 1.22 (0.68;2.2)      & 0.301                & 1.29 (0.85;1.94)     & 0.210                 & 0.87 (0.5;1.52)      & 0.285                \\
\textbf{CBPS}   & 1.26 (0.70;2.26)      & 0.300                  & 1.33 (0.88;2.02)     & 0.213                & 0.77 (0.44;1.35)     & 0.283                \\
\textbf{SBW}    & -                     & -                  & 1.31 (0.87;1.99)     & 0.212                & -     & -                \\
\textbf{EBAL}   & 1.25 (0.70;2.26)      & 0.300                  & 1.33 (0.88;2.02)     & 0.213                & 0.77 (0.44;1.34)     & 0.283                \\
\textbf{PSM}    & 0.96 (0.19;4.73)     & 0.817                & 0.57 (0.14;2.39)     & 0.731                & 0.57 (0.14;2.39)     & 0.731                \\
\textbf{OM}     & 1.24 (0.66;2.3)      & 0.318                & 1.35 (0.91;2.02)     & 0.203                & 0.93 (0.55;1.58)     & 0.271                \\
\textbf{Naive}  & 1.23 (0.70;2.18)      & 0.291                & 1.17 (0.79;1.72)     & 0.198                & 0.82 (0.48;1.38)     & 0.267                \\ \hline
\end{tabular}
    \begin{tablenotes}
      \small
      \item \textit{First column}: Method implemented, ROW: Robust Optimal Weights; IPW: Inverse Probability Weighting (propensity scores were estimated with SuperLearner with linear  regression  model  with  only  main  effects,  and random  forest in the library of algorithms); BalSL: Balance SuperLearner; GBM: propensity scores were estimated with Gradient Boosting Machine; CBPS: Covariate Balancing Propensity Score; SBW: Stable Balancing Weights; EBAL: Entropy Balancing; PSM: Propensity Score Matching;  OM: (outcome model) Cox  proportional  hazard model including confounders and treatment; Naive:  Cox  proportional  hazard model including only the treatment. \textit{Second, Fifth and Eight columns}: Minimum absolute standardized mean difference between confounders and treatment across 33 confounders. \textit{Third, Sixth and Ninth columns}: Median absolute standardized mean difference between confounders and treatment across 34 confounders. \textit{Fourth, Seventh and Tenth columns}: Maximum absolute standardized mean difference between confounders and treatment across 33 confounders.
    \end{tablenotes}
  \end{threeparttable}
\end{table}

\subsection{The effect of red meat consumption on colon cancer}
\label{redmeat_colon}

Colon cancer is the third most common cause of cancer-related death in the United States. Although epidemiological studies have shown that excess consumption of red meat may be related to colon cancer \citep{larsson2005red,larsson2006meat}, its consumption in the United States has not been decreasing in the past few decades \citep{zeng2019trends}. 

In this section,  using data from the WHI observational study, we aim at evaluating the effect of red meat consumption on time to colon cancer among postmenopausal women aged 50-79 years. Following \cite{song2004prospective}, we defined red meat as the sum of beef, hamburger, lamb or pork as a main dish or a sandwich or mixed dish, and all processed red meat. We defined consumption as medium servings per day of red meat. In addition to the 34 confounders described in Section \ref{ht_chd}, we also consider the time since menopause (numeric) and the use or not of estrogen plus progestin therapy (yes, no) as additional confounders.

\subsubsection{Models setup}

We obtained ROW by solving optimization problem \eqref{op1}-\eqref{op4} setting $\delta=0.001$ in constraint \eqref{op2}. To obtain the set of ROW we used the \texttt{R} interface of \texttt{Gurobi}.  We compared ROW with IPW in which we estimated the generalized propensity scores by using SuperLearner with the following library of algorithms: linear  regression  model  with  only  main  effects,  and random  forest; BalSL with the same library of algorithms as for IPW; GBM (with mean Pearson correlation between covariates and treatment as stopping method, interaction depth equal to 4, number of trees equal to 20,000, shrinkage equal to 0.0005 and bag fraction equal to 1); CBPS and npCBPS containing only the covariate balancing conditions, and OM, a Cox proportional hazard model including confounders and red meat consumption. As suggested by \cite{robins2000marginal}, to compute the generalized propensity scores for IPW, BalSL and GBM we assumed the conditional density of the treatment to be Normal.  Once the sets of weights were obtained, we plugged the weights into a weighted Cox regression regressing red meat consumption on the time to colon cancer. We computed robust standard errors. 

\subsubsection{Results}

\begin{figure}
\begin{center}
\includegraphics[scale=.55]{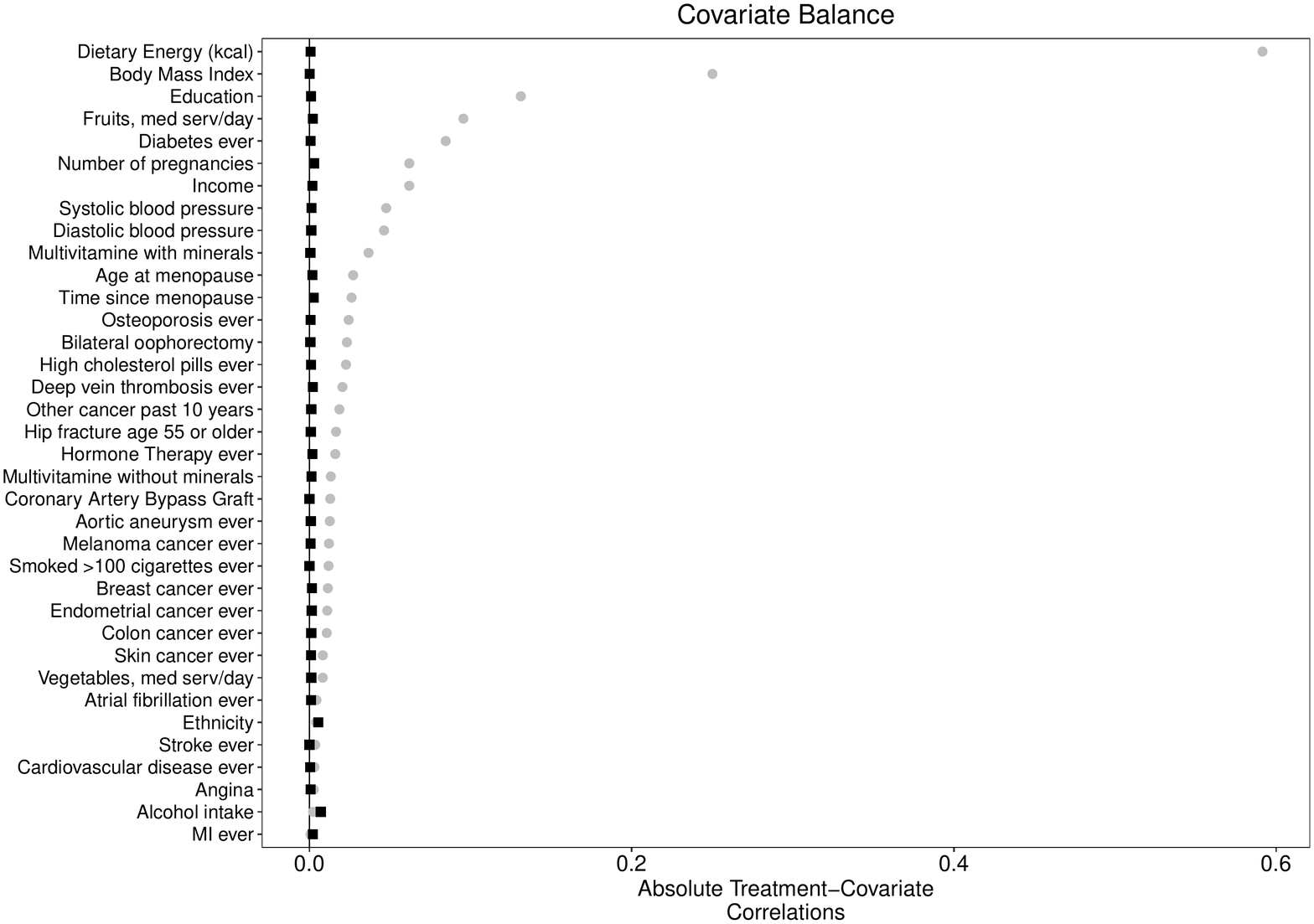}
\end{center}
\caption{Adjusted (weighted by ROW) (black squares) and unadjusted (grey dots) absolute correlation between confounders and treatment (red meat consumption). 
\label{fig_corr_cont} }
\end{figure}

\textbf{Covariate balance.} The first, second and third columns of Table \ref{table_c} shows the minimum, median and maximum absolute correlation between confounders and red meat consumption across 36 confounders. ROW resulted in the lowest maximum absolute correlation, followed by npCBPS with a maximum absolute correlation more than 20 times higher than that of ROW. In addition, ROW required only 12 seconds to find a solution, compared with much higher computational times needed by the other methods (fourth column of Table \ref{table_c}). Figure \ref{fig_corr_cont} shows absolute correlations between the 36 confounders and red meat consumption. As shown in Table \ref{table_c}, ROW resulted in low absolute correlations across all confounders 

\noindent
\textbf{Outcome analysis.}
The last two columns of Table \ref{table_c} show the estimated marginal hazard ratio, its 95\%  confidence interval and robust standard error of the impact of red meat consumption on time to colon cancer. Proportional hazards tests \citep{grambsch1994proportional} resulted in p-values greater than 0.05 for all models.  Based on assumptions 2.1-2.3 of Section \ref{sec:notation_assumptions}, our simulations results presented in Section \ref{sec:simu}, and the absolute correlations showed in Figure \ref{fig_corr_cont}, we conclude that red meat consumption is statistically associated with higher risk of colon cancer among $n=24,069$ postmenopausal women aged 50-79 years enrollend in the WHI observational study (September 1993-September 2010), \ie, $\hat{HR}$ and $95\%$ confidence intervals equal to  1.68 (1.19;2.37).

\begin{table}[]
\centering
\begin{threeparttable}
\caption{Minimum, median and maximum absolute correlation between confounders and red meat consumption across 36 confounders, computational time, marginal hazard ratio estimate, 95\% confidence intervals and robust standard error of the effect of red meat consumption on time to colon cancer among postmenopausal  women between 50 and 79 - WHI observational study, 1993-2010, $n=24,069$. \label{table_c}}
\begin{tabular}{ccccccc}
\hline
\textbf{}       & \multicolumn{3}{c}{\textbf{Absolute Correlation}} & \multicolumn{1}{c}{\textbf{Time}}   & \multicolumn{1}{c}{} &                                                          \\
\textbf{Method} & \multicolumn{1}{c}{\textbf{Min}} & \multicolumn{1}{c}{\textbf{Median}} & \multicolumn{1}{c}{\textbf{Max}} & \multicolumn{1}{c}{\textbf{sec}} & \multicolumn{1}{c}{$\hat{HR}$ $(95\% \text{CI})$} & SE\\ \hline
\textbf{ROW}    & \textless{}0.001                 & 0.0011                              & 0.0070  & 12.0  & 1.68 (1.19;2.37)    & 0.17                          \\
\textbf{IPW}    & \textless{}0.001                 & 0.0097                              & 0.2816    & 366.5    & 1.26 (1.05;1.5)     & 0.09                   \\
\textbf{BalSL}  & \textless{}0.001                 & 0.0101                              & 0.2309    & 367.2    & 1.28 (1.06;1.54)  & 0.09                   \\
\textbf{GBM}    & \textless{}0.001                 & 0.0163                              & 0.5924 & 778.3   & 1.17 (1.00;1.37)  & 0.08                       \\
\textbf{CBPS}   & \textless{}0.001                 & 0.0148                              & 0.3489  & 64.5 & 1.23 (0.64;2.36)  & 0.33                        \\
\textbf{npCBPS} & \textless{}0.001                 & 0.0094                              & 0.1523 & 9724.1  & 1.22 (0.98;1.52)  & 0.11                         \\
\textbf{OM}     & -                                & -                                   & -       & 0.2          & 1.28 (1.00;1.63)  & 0.13                \\
\textbf{Naive}  & \textless{}0.001                 & 0.0163                              & 0.5915   & 0.1    & 1.17 (0.99;1.39)  & 0.08                    \\ \hline
\end{tabular}
    \begin{tablenotes}
      \small
      \item \textit{First column}: Method implemented, ROW: Robust Optimal Weights; IPW: Inverse Probability Weighting (propensity scores were estimated with SuperLearner with linear  regression  model  with  only  main  effects,  and random  forest in the library of algorithms); BalSL: Balance SuperLearner; CBPS: Covariate Balancing Propensity Score; npCBPS: non-parametric CBPS; OM: Cox  proportional  hazard model - outcome model.  \textit{Second to fourth columns}: minimum, median and maximum absolute correlation across confounders. \textit{Fifth column}: Time in seconds.   \textit{Sixth column}: Marginal hazard ratio estimate and 95\% confidence intervals (computed with robust standard errors). \textit{Seventh column}:  robust standard error. 
    \end{tablenotes}
  \end{threeparttable}
\end{table}

\section{Conclusions}
\label{sec:conclusions}

Unbiased estimation of the effect of binary and continuous treatments using observational data is crucial for medical decision making. In this paper, we introduced a method based on the framework of \cite{yiu2018covariate} and on a convex constrained quadratic optimization problem that finds weights with minimal variance, thus controlling for extreme weights, while satisfying constraints on the sample correlation between confounders and treatment, thus targeting covariate balance. ROW performed well across levels of practical positivity violation, misspecification and censoring for both binary and continuous treatments. In addition, in this paper we have shown that methods that target covariate balance, like ROW, CBPS, SBW and EBAL perfom well in terms of absolute bias and root mean squared error compared with propensity score methods, like IPW, regardless of what method is used to estimate the propensity scores.

In addition to the hazard ratio, which use for estimating treatment effects has been discouraged \citep{bartlett2020hazards,hernan2010hazards}, ROW can be used to estimate other survival causal parameters of interest like the contrasts of survival probabilities or of specified quantiles and differences or ratios of restricted survival means \citep{mao2018propensity,bartlett2020hazards}. Also, ROW can be used to estimate effects of binary and continuous treatments on binary outcomes, (marginal odds ratio), and continuous outcomes (average and quantile treatment effects). ROW can also be use to estimate effects of multi-value treatments. To do so, if the interest is to make inference on each of the treatment's contrasts, we suggest to run pair-wise comparisons by applying ROW as if it was a binary treatment. Alternatively, one can consider the multi-value treatment as a continuous treatment (as suggested by \cite{fong2018covariate}). Future research is needed to thoroughly evaluate the performance of ROW with multi-value treatments. 

As presented in our simulation and in our case studies, we suggest to considered categorical variables (such as education for instance) as numeric, and balance them accordingly. In addition, in our simulations and case studies we only considered balancing linear covariates. Quadratic or higher order and interaction terms can be balanced by adding them into constraint \eqref{op2}. If no solution to optimization problem \eqref{of1}-\eqref{op4} exists, we suggest to increase the parameter $\delta$ of \eqref{op2} and re-run the solver. We also suggest to evaluate covariate balance, defined as the absolute standardized mean difference for a binary outcome and as the absolute correlation for a continuous outcome for each new set of ROW. 




\newpage
\bibliographystyle{chicago}
\bibliography{row}

\newpage
\bigskip
\begin{center}
{\large\bf SUPPLEMENTARY MATERIAL}
\end{center}

\section{Simulations}
\label{sm:simu}

In this section, we provide additional simulations' results evaluating 1) the naive (the inverse of the observed Fisher information of the coefficient), robust and bootstrap standard error estimator \ref{sm:simu_se_cove}; 2) the impact of practical positivity violation, misspecification and censoring on coverage of 95\% confidence intervals \ref{sm:simu_se_cove}; 3) the impact of sample sizes \ref{sm:simu_samplesize} and 4) the impact of the number of covariates included in the analysis \ref{sm:simu_num_cove}, on absolute bias, root mean squared error, balance and computational time in seconds.

\subsection{Standard error and coverage}
\label{sm:simu_se_cove}

\begin{figure}
\begin{center}
\includegraphics[scale=0.6]{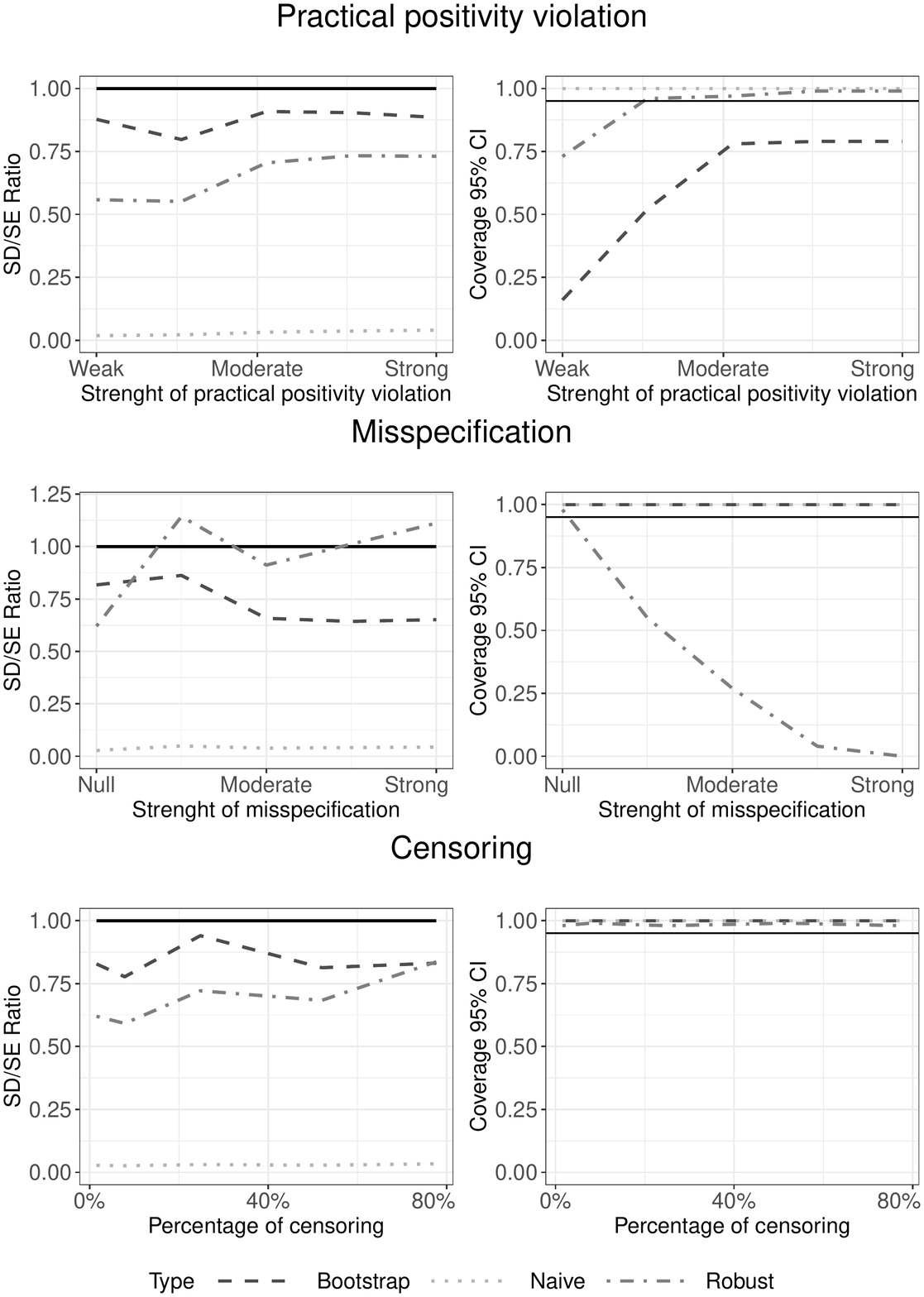}
\end{center}
\caption{\footnotesize \textit{Left panels}: ratios between the standard deviation of the estimated hazard ratio across simulations and the bootstrap (dashed), naive (the inverse of the observed Fisher information of the coefficient)(dotted) and robust (dotted-dashed) standard errors across levels of practical positivity violation (top panels), misspecification (middle panels) and censoring (bottom panels) for the binary treatment. \textit{Right panels}: coverage of the 95\% confidence interval using the bootstrap (with normal approximation confidence intervals)(dashed), naive (dotted) and robust (dotted-dashed) standard errors across levels of practical positivity violation (top panels), misspecification (middle panels) and censoring (bottom panels) for the binary treatment.
\label{fig_se_cove} }
\end{figure}

\begin{figure}
\begin{center}
\includegraphics[scale=0.6]{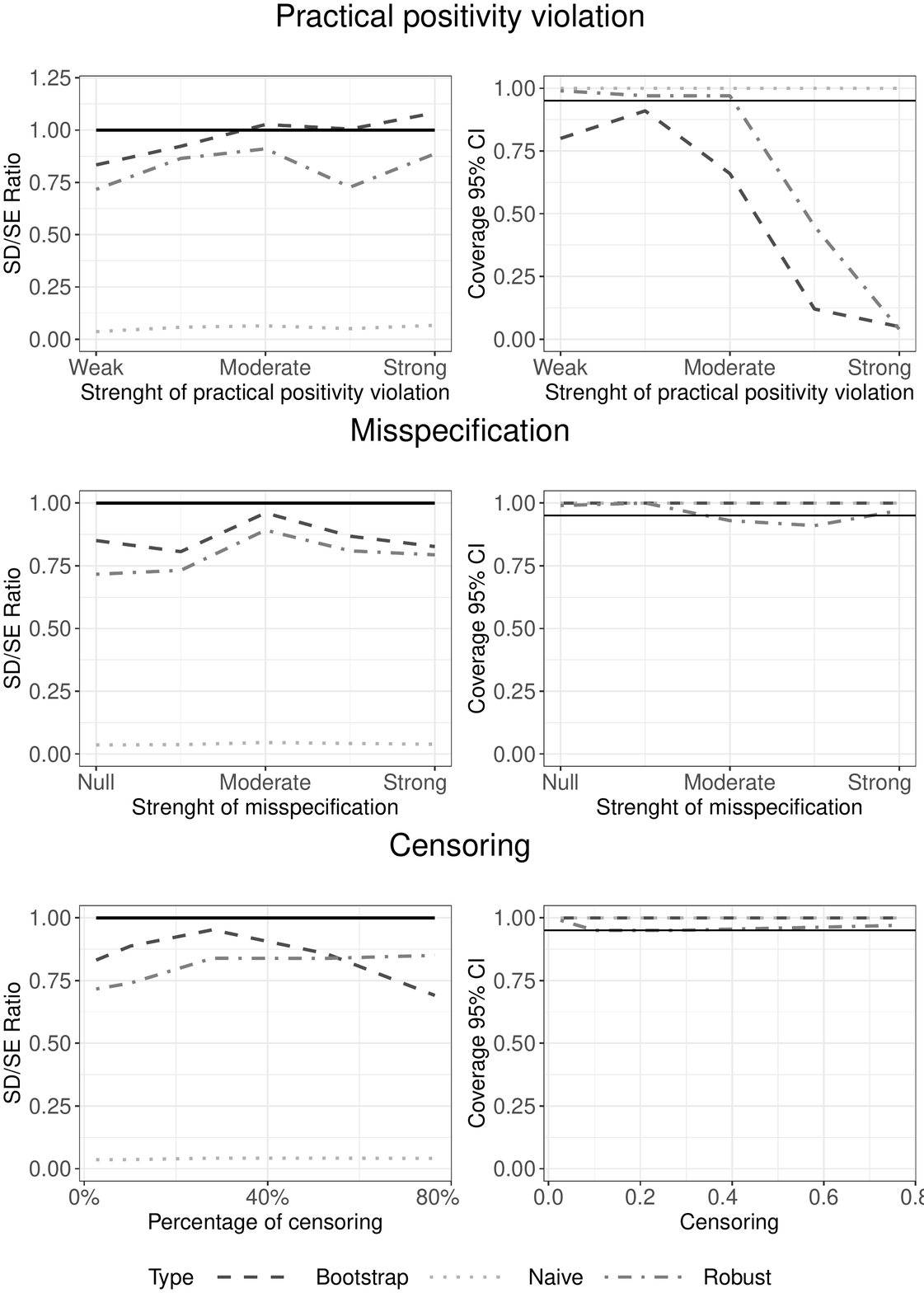}
\end{center}
\caption{\footnotesize \textit{Left panels}: ratios between the standard deviation of the estimated hazard ratio across simulations and the bootstrap (dashed), naive (the inverse of the observed Fisher information of the coefficient)(dotted) and robust (dotted-dashed) standard errors across levels of practical positivity violation (top panels), misspecification (middle panels) and censoring (bottom panels) for the continuous treatment. \textit{Right panels}: coverage of the 95\% confidence interval using the bootstrap (with normal approximation confidence intervals)(dashed), naive (dotted) and robust (dotted-dashed) standard errors across levels of practical positivity violation (top panels), misspecification (middle panels) and censoring (bottom panels) for the continuous treatment.
\label{fig_se_cove_c} }
\end{figure}

We considered the same simulation scenario as that described in Section \ref{sec:simu} of the original manuscript. We used non-parametric bootstrap with normal approximation confidence intervals \citep{davison1997bootstrap}. Left panels of Figures \ref{fig_se_cove} and \ref{fig_se_cove_c} show the ratios between the standard deviation of the estimated hazard ratio across simulations and the bootstrap, robust and naive standard errors for the binary treatment (Figure \ref{fig_se_cove}) and for the continuous treatment (Figure \ref{fig_se_cove_c})  scenarios. The Naive estimator resulted in higher standard error compared with the empirical standard deviation (lower values of the ratio) and consequential overcoverage across all levels of practical positivity violation, misspecification and censoring for both binary and continuous treatments. Robust and bootstrap standard errors behaved similarly across most levels of practical positivity violation, misspecification and censoring and type of treatments. Under strong misspecification in the binary treatment scenario (middle panel of Figure \ref{fig_se_cove}), ROW using the robust standard error estimator substantially undercovered the 95\% confidence interval, while it overcovered it while using bootstrap and the naive standard error. Overall, robust and bootstrap standard errors resulted in slightly higher standard errors compared with the standard deviation of the estimated hazard ratio, and consequently resulted in higher coverage. As expected, when standard errors are close to the standard deviation of the estimated hazard ratio and ROW has bias, ROW exhibits undercoverage (top right panel of Figure \ref{fig_se_cove_c}).

\subsection{Sample size}
\label{sm:simu_samplesize}

\begin{figure}
\begin{center}
\includegraphics[scale=0.55]{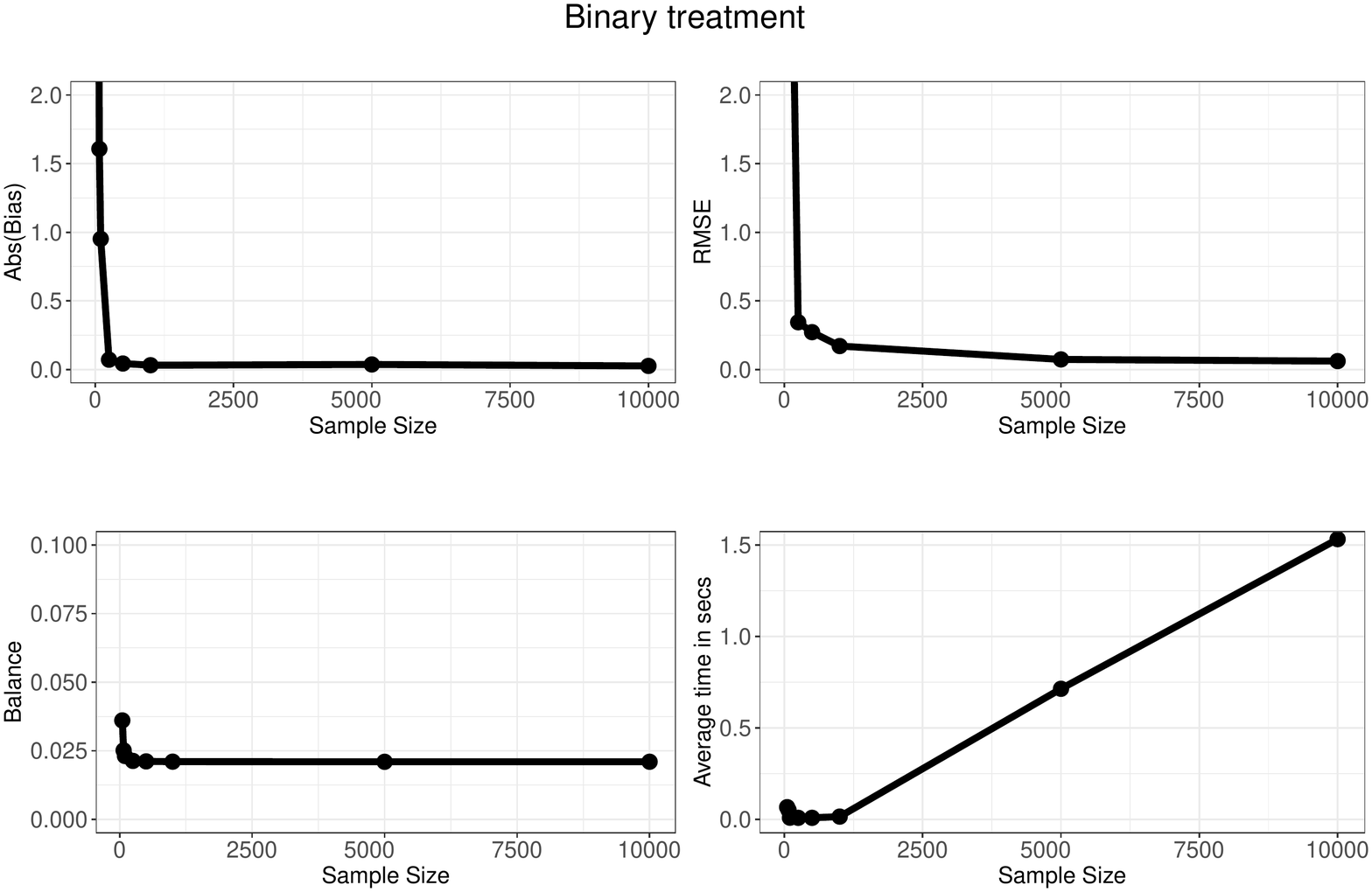}
\end{center}
\caption{\footnotesize Absolute bias (top left panel), root mean squared error (top right panel), absolute standardized mean difference across four confounders (left bottom panel)  and the average computational time in seconds to obtain a solution (right bottom panel) across levels of sample sizes for the binary treatment scenario. 
\label{fig_ss_b} }
\end{figure}

We considered sample sizes equal to $50, 75, 100, 250, 500, 5,000,$ and $10,000$. We computed the expected survival time $t$ as described in Section \ref{simu_setup} of the original manuscript, where $X_{1:4} \sim \text{MultiN}(\mu, \Sigma)$, $\mu=(0,0,0,0)$ and $\Sigma=I_4$, where $I_4$ is the identity matrix. We considered $\beta=1.4$ (the coefficients for the confounders in the outcome model - Section \ref{simu_meth}) and $\gamma=1.5$ (the coefficients of the confounders in the treatment model - Section \ref{simu_bin_cont}) for all four confounders. To reflect only the impact of sample size, we considered a moderate practical positivity violation scenario ($\gamma=1$ and $\nu=0.6$ as in Section \ref{simu_ppv}), a null misspecification ($\tau=0$ as in Section \ref{simu_meth}) and a low percentage of censored observation ($\epsilon=0$ as in Section \ref{simu_cens}). Figure \ref{fig_ss_b} shows the absolute bias (top lef panel), the root mean squared error (RMSE) (top right panel), absolute standardized mean difference across the four confounders (Balance) (left bottom panel)  and the average computational time in seconds to obtain a solution (right bottom panel) across levels of sample sizes for the binary treatment scenario. Figure \ref{fig_ss_b} shows those for the continuous treatment.  For both treatments, absolute bias and RMSE decreased with increasing sample sizes. Balance was kept low across all sample sizes. The computational time in seconds increased up to 1.5 seconds for $n=10,000$.

\begin{figure}
\begin{center}
\includegraphics[scale=0.5]{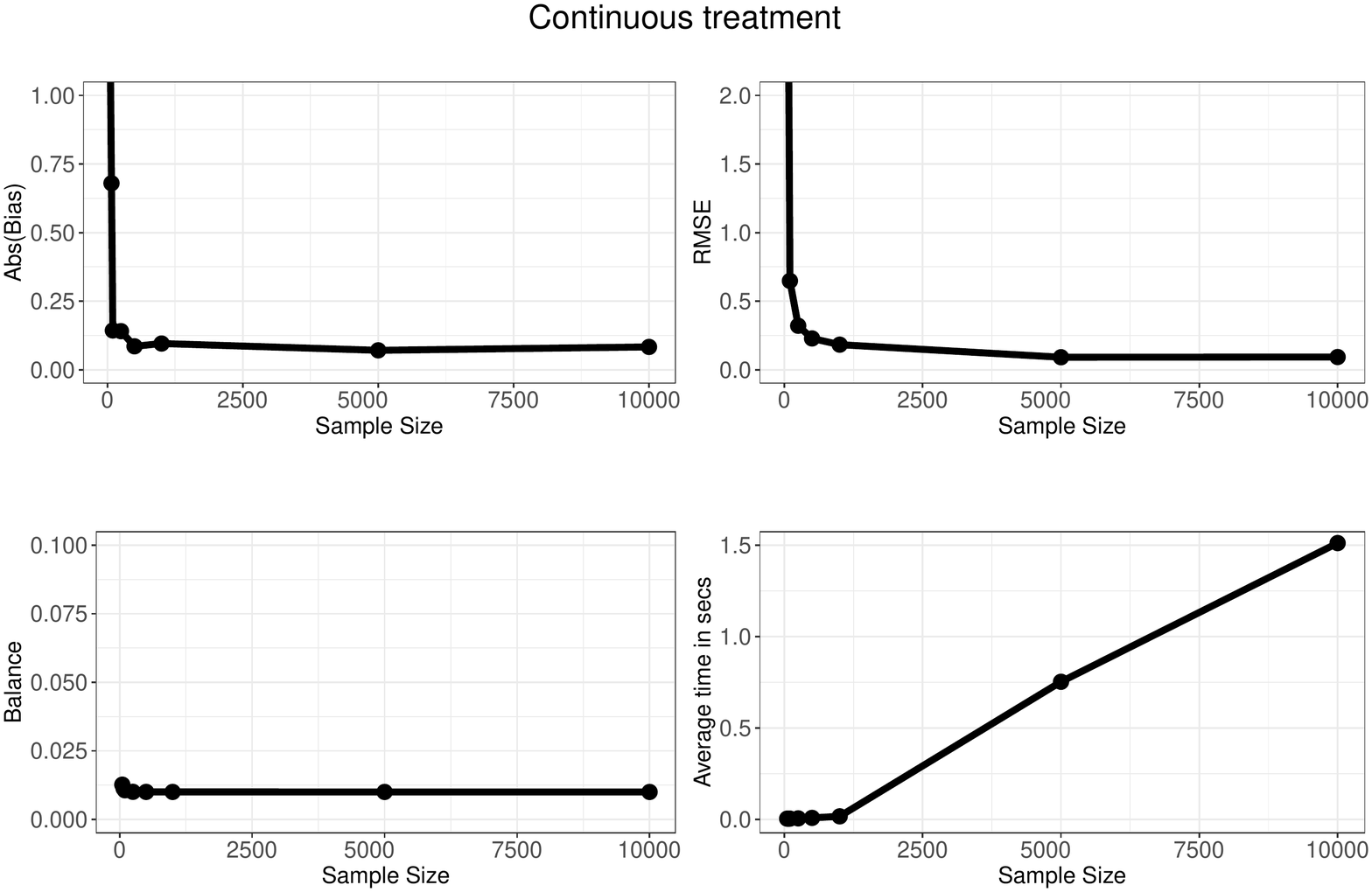}
\end{center}
\caption{\footnotesize Absolute bias (top left panel), root mean squared error (top right panel), absolute standardized mean difference across four confounders (left bottom panel)  and the average computational time in seconds to obtain a solution (right bottom panel) across levels of sample sizes for the continuous treatment scenario. 
\label{fig_ss_c} }
\end{figure}

\subsection{Number of confounders}
\label{sm:simu_num_cove}

\begin{figure}
\begin{center}
\includegraphics[scale=0.5]{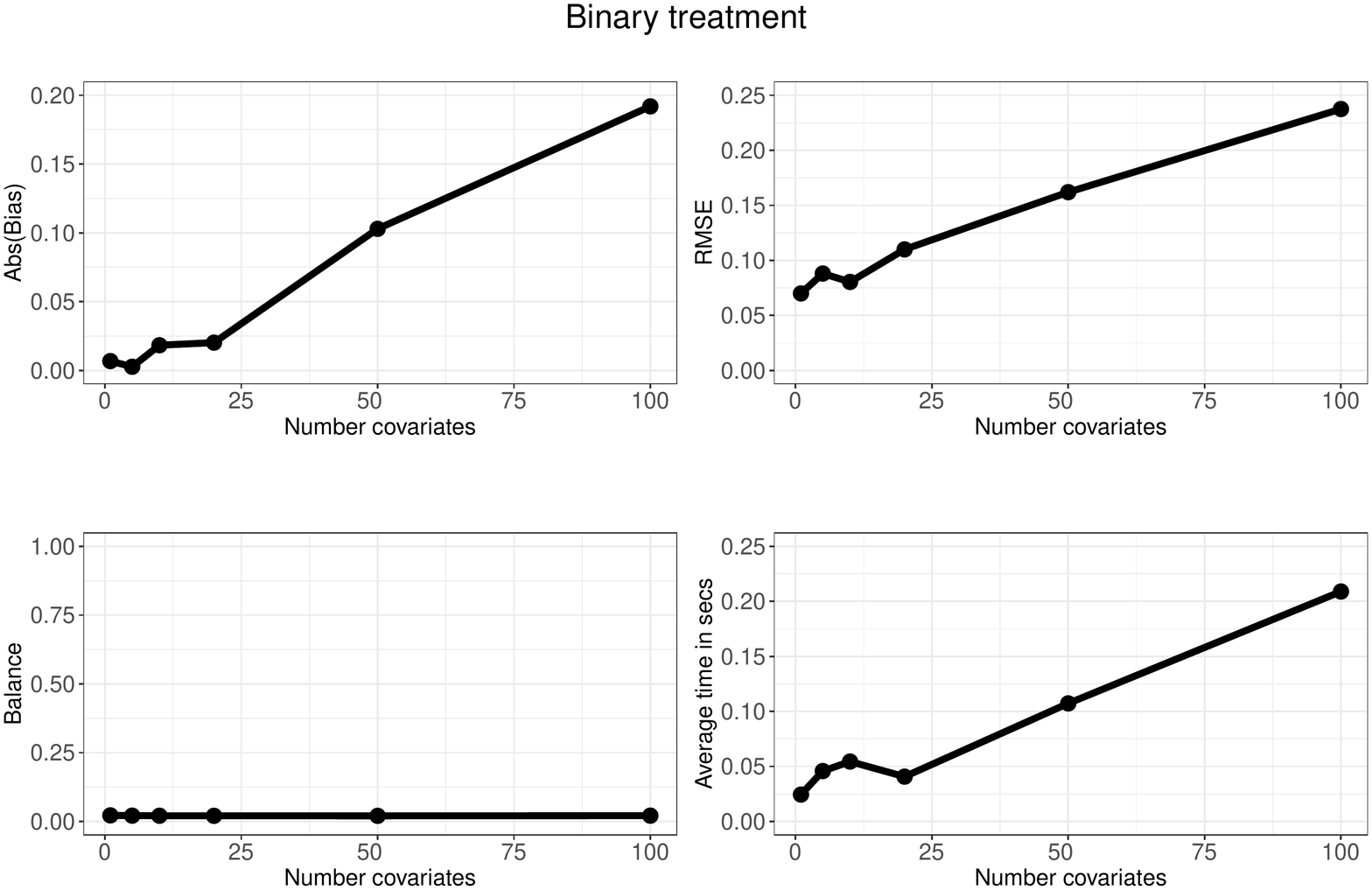}
\end{center}
\caption{\footnotesize Absolute bias (top left panel), root mean squared error (top right panel), absolute standardized mean difference across four confounders (left bottom panel)  and the average computational time in seconds to obtain a solution (right bottom panel) across number of confounders for the binary treatment scenario. 
\label{fig_num_c_b} }
\end{figure}

\begin{figure}
\begin{center}
\includegraphics[scale=0.5]{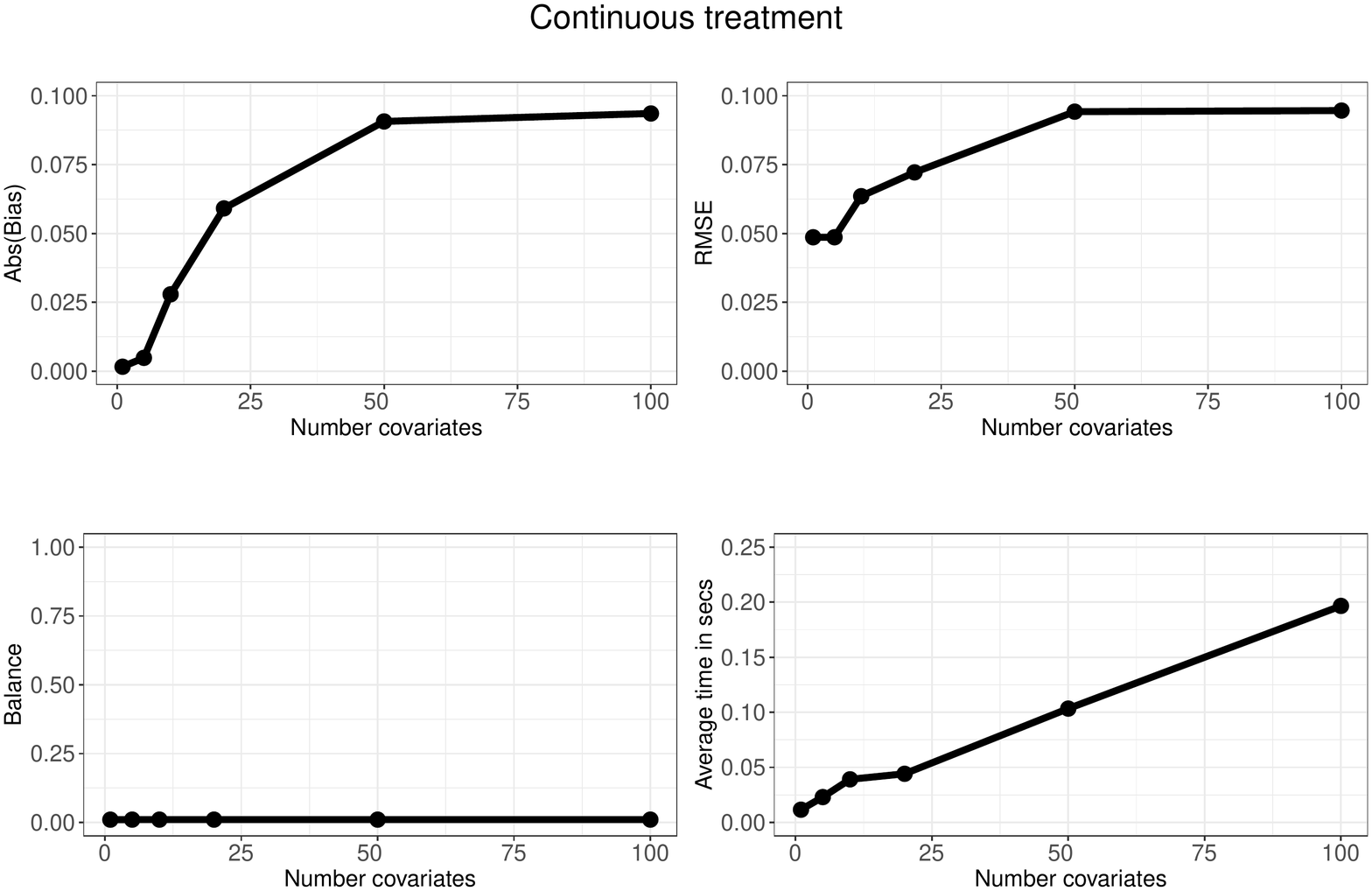}
\end{center}
\caption{\footnotesize Absolute bias (top left panel), root mean squared error (top right panel), absolute standardized mean difference across four confounders (left bottom panel)  and the average computational time in seconds to obtain a solution (right bottom panel) across number of confounders for the continuous treatment scenario. 
\label{fig_num_c_c} }
\end{figure}

We considered the following number of confounders $1, 5, 10, 20, 50,$ and $100$ and $n=1,000$.  We computed the expected survival time $t$ as described in Section \ref{simu_setup} of the original manuscript, where $X_{1:K} \sim \text{MultiN}(\mu, \Sigma)$, $\mu=$\textbf{0} and $\Sigma=I_K$, where, \textbf{0} and $I_K$ are the vector of zero of dimension $K$, and the identity matrix of dimension $K \times K$, where $K$ is total number of confounders. We considered $\beta=0.1$ (the coefficients for the confounders in the outcome model - Section \ref{simu_meth}) and $\gamma=0.1$ (the coefficients of the confounders in the treatment model - Section \ref{simu_bin_cont}) for all confounders. As for the evaluation of the impact of sample sizes, we considered a moderate practical positivity violation scenario ($\gamma=1$ and $\nu=0.6$ as in Section \ref{simu_ppv}), a null misspecification ($\tau=0$ as in Section \ref{simu_meth}) and a low percentage of censored observation ($\epsilon=0$ as in Section \ref{simu_cens}). Figure \ref{fig_num_c_b} shows the absolute bias (top lef panel), the root mean squared error (RMSE) (top right panel), absolute standardized mean difference across the four confounders (Balance) (left bottom panel)  and the average computational time in seconds to obtain a solution (right bottom panel) across numbers of confounders for the binary treatment scenario. Figure \ref{fig_num_c_c} shows those for the continuous treatment.  For both treatments, absolute bias and RMSE increased with increasing number of confounders. Balance was kept low across all levels. The computational time in seconds slightly increased from below 0.05 seconds to 0.20 seconds (100 confounders).

\section{Case studies}
\label{sm:sec:case-studies}

In this Section, we provide additional results of the case studies presented in Section \ref{sec:case-studies} in the original manuscript. 

\begin{table}[]
\centering
\begin{threeparttable}
\caption{Computational time in seconds needed to obtain a solution across categories of time since menopause (0-10; 11-20; 20+ years), WHI observational study, $n=24,069$, 1993-2010. \label{table_b3}}
\begin{tabular}{cccc}
\hline
\textbf{}                           & \multicolumn{3}{c}{\textbf{Time since menopause}}                  \\ \cline{2-4} 
\textbf{}                           & \textbf{0-10 Years}  & \textbf{11-20 Years} & \textbf{20+ Years}   \\
\textbf{}                           & \textbf{$n=6,661$}  & \textbf{$n=9,592$} & \textbf{$n=7,816$}   \\
                                    & \textbf{Time (sec)}  & \textbf{Time (sec)}  & \textbf{Time (sec)}  \\ \cline{2-4} 
\multicolumn{1}{l}{\textbf{Method}} & \multicolumn{1}{l}{} & \multicolumn{1}{l}{} & \multicolumn{1}{l}{} \\ \cline{1-1}
\textbf{ROW}                        & 1.0                  & 2.8                  & 1.6                \\
\textbf{IPW}                        & 218.4                & 332.5                & 238.1                \\
\textbf{BalSL}                      & 213.6                & 312.2                & 225.6                \\
\textbf{GBM}                        & 83.7                 & 115.2                & 93.3                 \\
\textbf{CBPS}                       & 4.6                  & 7.9                  & 6.2                  \\
\textbf{SBW}                        & 653.5$^\ast$                    & 2310.8                  & 1313.8$^\ast$                  \\
\textbf{EBAL}                       & 1.1                  & 1.6                  & 1.3                  \\
\textbf{PSM}                        & 236.0                & 338.1                & 229.1                \\
\textbf{OM}                         & 0.1                  & 0.1                  & 0.1                  \\
\textbf{Naive}                      & \textless{}0.1       & \textless{}0.1       & \textless{}0.1       \\ \hline
\end{tabular}
    \begin{tablenotes}
      \small
      \item \footnotesize \textit{First column}: Method implemented, ROW: Robust Optimal Weights; IPW: Inverse Probability Weighting (propensity scores were estimated with SuperLearner with linear  regression  model  with  only  main  effects,  and random  forest in the library of algorithms); BalSL: Balance SuperLearner; GBM: propensity scores were estimated with Gradient Boosting Machine; CBPS: Covariate Balancing Propensity Score; SBW: Stable Balancing Weights; EBAL: Entropy Balancing; PSM: Propensity Score Matching;  OM: (outcome model) Cox  proportional  hazard model including confounders and treatment; Naive:  Cox  proportional  hazard model including only the treatment. \textit{Second, Third and Fourth columns}: Computational time in seconds and sample size within each category of time since menopause.  $^\ast$ Time at which SBW stopped without finding a solution with balance tolerance set equal to 0.0001.
    \end{tablenotes}
  \end{threeparttable}
\end{table}

\begin{figure}
\begin{center}
\includegraphics[scale=0.6]{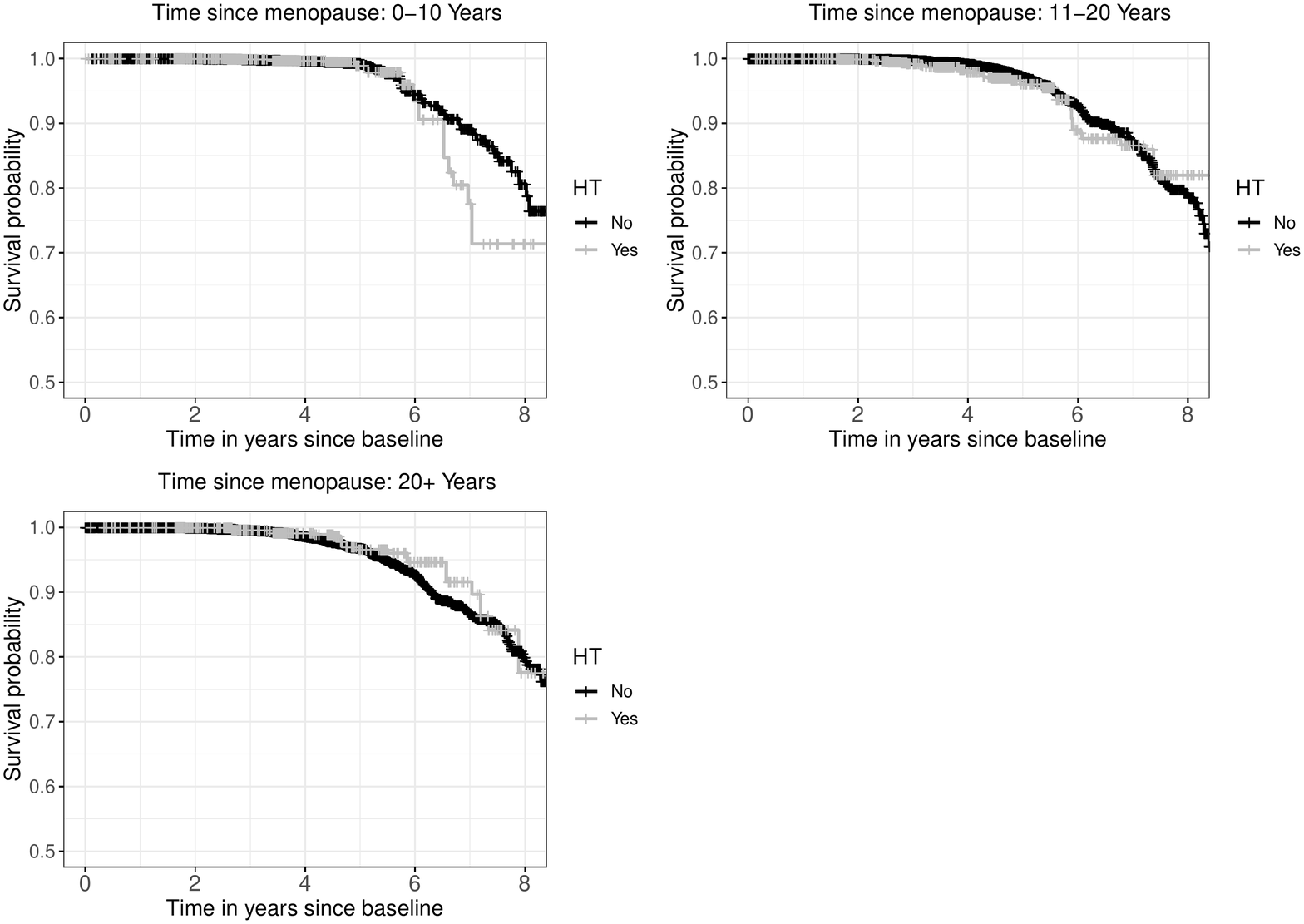}
\end{center}
\caption{\footnotesize Kaplan-Meier curves weighted by ROW for estrogen plus progestin (HT; Yes versus No) across categories of time since menopause.
\label{fig_km} }
\end{figure}

\section{Additional results}
\label{sm:sec:additional results}

In this Section, we investigate the distribution of ROW across different treatment, covariate relationships. We considered eight scenarios:

\begin{enumerate}
    \item \textit{Linear dependence}: the treatment, $A$ has a linear dependence on the confounder, $X$ as 
    \begin{equation*}
        X\sim N(0,1) \quad \text{and} \quad A\sim X+N(0,1)
    \end{equation*}
    
    \item \textit{Nonlinear dependence quadratic}: the treatment, $A$ has a quadratic dependence on the confounder, $X$ as 
    \begin{equation*}
        X\sim N(0,1) \quad \text{and} \quad A\sim X + X^2+N(0,1)
    \end{equation*}
    
        \item \textit{Nonlinear dependence cubic}: the treatment, $A$ has a cubic dependence on the confounder, $X$ as 
    \begin{equation*}
        X\sim N(0,1) \quad \text{and} \quad A\sim 0.5 (X+0.1)^3+N(0,1)
    \end{equation*}
    
    \item \textit{Nonlinear dependence without correlation}: the treatment, $A$ has a lattice-dependence on the confounder, $X$ as 
    \begin{equation*}
        X\sim Unif(-.5,.5) \quad \text{and} \quad A \sim \left\{
                \begin{array}{ll}
                  N(0,\frac{1}{3}) \quad &\text{if } X \leq \| \frac{1}{6} \| \\
                  \frac{1}{2}\left( N(1,\frac{1}{3}) + N(-1,\frac{1}{3}) \right) \quad &\text{otherwise} 
                \end{array}
              \right. 
    \end{equation*}
    
    \item \textit{Sinusoidal dependence}: the treatment, $A$ has a sinusoidal dependence on the confounder, $X$ as 
    \begin{equation*}
        X\sim N(0,4) \quad \text{and} \quad A\sim \sin{X}+N(0,0.1)
    \end{equation*}

    \item \textit{Independence}: the treatment, $A$ is independent on the confounder, $X$ as 
    \begin{equation*}
        X \sim N(0,1) \quad \text{and} \quad A \sim N(0,1)
    \end{equation*}

        \item \textit{Right-skewed}: the treatment, $A$ is right-skewed and depends on the confounder, $X$ as 
    \begin{equation*}
        X \sim Beta(1,5) \quad \text{and} \quad A \sim 4X + LogN(0,0.7)
    \end{equation*}
    
            \item \textit{Left-skewed}: the treatment, $A$ is left-skewed and depends on the confounder, $X$ as 
    \begin{equation*}
        X \sim Beta(5,1) \quad \text{and} \quad A \sim 4X + Beta(5,1)
    \end{equation*}
    
\end{enumerate}

\begin{figure}
\begin{center}
\includegraphics[scale=0.6]{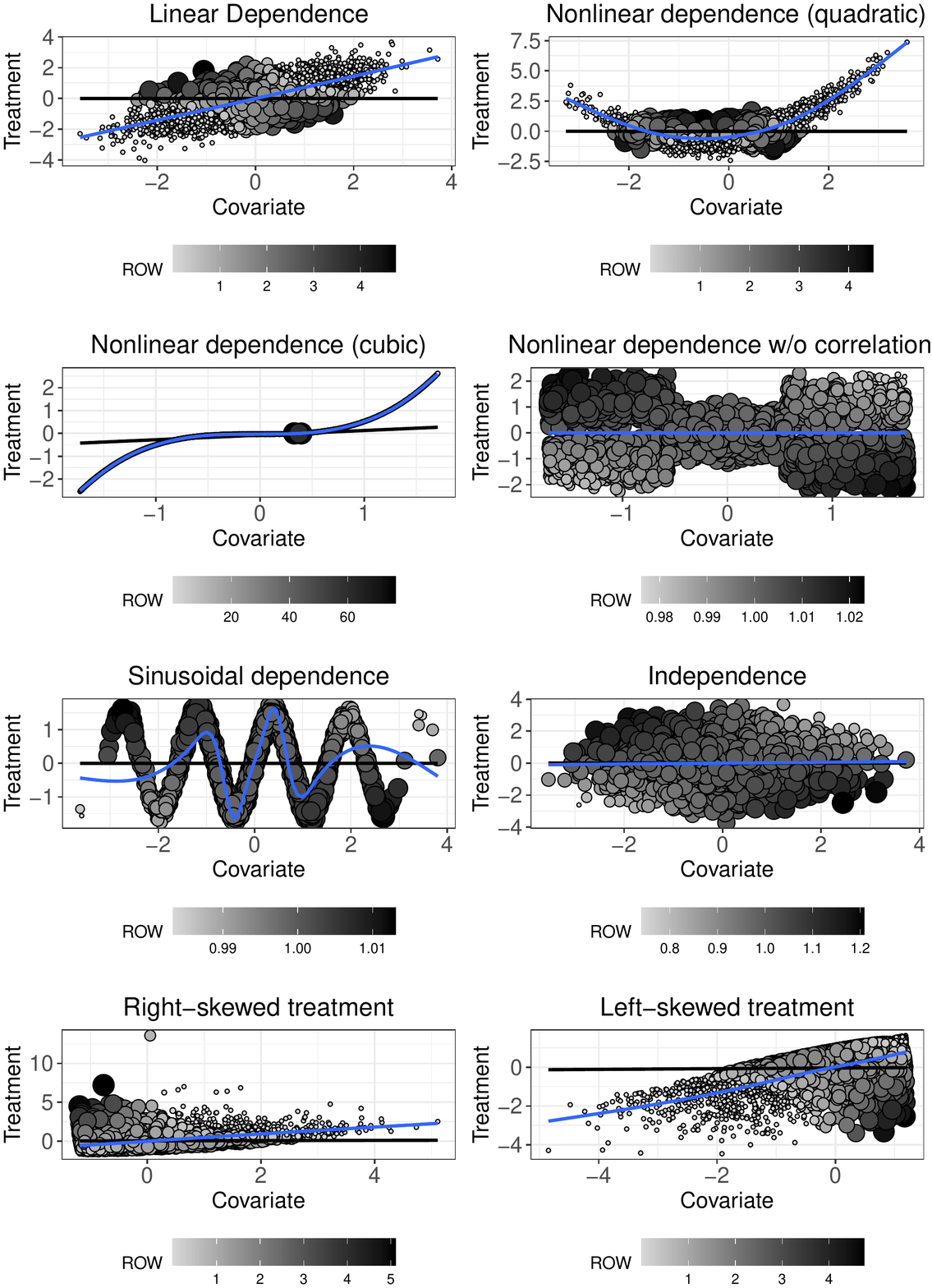}
\end{center}
\caption{\footnotesize Graphical representation of ROW balancing a covariate (x-axis) and continuous treatment (y-axis) across different covariate-treatment relationships.  Blue lines represent the true relationships between thebinary (a probit model) and the continuous (simple regression with normal errors and positive coefficient)treatment.   Black  lines  represent  the  relationship  between  treatments  and  covariate  after  weighting  for ROW. Size and color of the circles represent the individual weight assigned (the larger/darker the higher).
\label{fig_deps} }
\end{figure}

Figure \ref{fig_deps} shows the relationships before (blue lines) and after (black lines) weighting for ROW across different covariate-treatment relationships. ROW was almost uniformly distributed under nonlinear dependence without correlation, sinusoidal and under independence. ROW presented larger weights under nonlinear cubic dependence. We balanced linear and terms for obtaining ROW under the quadratic scenario. We balanced linear, quadratic and cubic terms for obtaining ROW under the cubic scenario. Regardeless of the rigth or left skeweness of the treatment, ROW successfully eliminated the relationship between the covariate and the treatment. ROW could be consequentely used when continuous treatments are skweded, such as for example, when interested in evaluating treatment doses, number of cigarettes smoked in one day, or daily food consumption. 

\end{document}